\documentclass[preprint]{aastex}

\newcommand{\oxyone}{O~{\footnotesize{I}}}       % O I
\newcommand{\oxysix}{O~{\footnotesize{VI}}}       % O VI
\newcommand{\nitfive}{N~{\footnotesize{V}}}       % N V
\newcommand{\carfour}{C~{\footnotesize{IV}}}      % C IV
\newcommand{\carthree}{C~{\footnotesize{III}}}    % C III
\newcommand{\cartwo}{C~{\footnotesize{II}}}       % C II
\newcommand{\carone}{C~{\footnotesize{I}}}        % C I
\newcommand{\hone}{H~{\footnotesize{I}}}          % H I
\newcommand{\arone}{Ar~{\footnotesize{I}}}        % Ar I
\newcommand{\sulsix}{S~{\footnotesize{VI}}}       % S VI
\newcommand{\fuse}{{\it FUSE}}       
\newcommand{\percc}{cm$^{-3}$}

\begin{document}

%============================================================================

\title{Observations of \oxysix\ Emission from the Diffuse Interstellar Medium}

\author{R. L. Shelton,\altaffilmark{1} 
J. W. Kruk,\altaffilmark{1}
E. M. Murphy,\altaffilmark{1}
B. G. Andersson,\altaffilmark{1} 
W. P. Blair,\altaffilmark{1} 
W. V. Dixon,\altaffilmark{2}
J. Edelstein,\altaffilmark{2} 
A.~W.~Fullerton,\altaffilmark{1} 
C. Gry,\altaffilmark{3, 4} 
J. C. Howk,\altaffilmark{1} 
E.~B.~Jenkins,\altaffilmark{5} 
J. L. Linsky,\altaffilmark{6} 
H. W. Moos,\altaffilmark{1} 
W. R. Oegerle,\altaffilmark{1} 
M. S. Oey,\altaffilmark{7} 
K. C. Roth,\altaffilmark{1}
D. J. Sahnow,\altaffilmark{1} 
R.~Sankrit,\altaffilmark{1}
B. D. Savage,\altaffilmark{8} 
K. R. Sembach,\altaffilmark{1}
J. M. Shull,\altaffilmark{6,9} 
O. H. W. Siegmund \altaffilmark{2}
A.~Vidal-Madjar,\altaffilmark{10} 
B. Y. Welsh,\altaffilmark{2} 
\& D. G. York\altaffilmark{11}}

\altaffiltext{1}{Department of Physics \& Astronomy, 
The Johns Hopkins University, Baltimore, MD 21218}
\altaffiltext{2} {Space Sciences Laboratory, University of California,
Berkeley, CA 94720-7450}
\altaffiltext{3} {ISO Data Center, European Space Agency, 
Astrophysics Division, P. O. Box
50727, 28080 Madrid, Spain}
\altaffiltext{4} {Laboratoir d'Astronomie Spatiale, B.P.8, 13376 Marseille 
cedex 12, France}
\altaffiltext{5}{Princeton University Observatory, Princeton, NJ 08544-1001}
\altaffiltext{6} {Center for Astrophysics and Space Astronomy and 
Department of Astrophysical and Planetary 
Sciences, University of Colorado, Boulder, CO 80309}
\altaffiltext{7} {Space Telescope Science Institute, 3700 San Martin
Drive, Baltimore, MD 21218}
\altaffiltext{8}{Department of Astronomy, University of Wisconsin-Madison, 
Madison, WI 53706}
\altaffiltext{9} {Joint Institute for Laboratory Astrophysics, 
University of Colorado and National Institute 
of Standards and Technology}
\altaffiltext{10}{Institut d'Astrophysique de Paris, 92 bis, Blvd. 
Arago, Paris 7504, France}
\altaffiltext{11} {Department of Astronomy and Astrophysics, University
of Chicago, Chicago, IL 60637}

\begin{abstract}

We report the first 
Far Ultraviolet Spectroscopic Explorer ({\it{FUSE}})
measurements of diffuse \oxysix\ ($\lambda\lambda$ 1032,1038)
emission from the general diffuse interstellar medium outside
of supernova remnants or superbubbles. 
We observed a $30\arcsec \times 30\arcsec$ region of the sky
centered at $l = 315\fdg0$ and $b = -41\fdg3$.
From the observed intensities
(2930$\pm$290(random)$\pm$410(systematic)  and 
1790$\pm$260(random)$\pm$250(systematic) 
photons cm$^{-2}$ s$^{-1}$ sr$^{-1}$ in the 
1032 and 1038~\AA\ emission lines, respectively), 
derived equations, and assumptions about the
source location, we calculate the intrinsic intensity,
electron density, thermal pressure, and emitting depth.
The intensities are too large for the emission to originate
solely in the Local Bubble.  Thus, we conclude that the Galactic
thick disk and lower halo also contribute.  High velocity
clouds are ruled out because there are none near the pointing
direction.  The calculated emitting depth is small, indicating
that the \oxysix-bearing gas fills a small volume.  
The observations can also be used to estimate the cooling rate 
of the hot interstellar medium and constrain models.  
The data also yield the
first intensity measurement of the
\cartwo\ 3s $^2$S$_{1/2}$ to 2p $^2$P$_{3/2}$ emission line 
at 1037~\AA\ and place upper limits
on the intensities of ultraviolet line emission from \carone,
\carthree, Si\ {\footnotesize{II}},
S\ {\footnotesize{III}}, S\ {\footnotesize{IV}}, S\ {\footnotesize{VI}},  
and Fe\ {\footnotesize{III}}.

\end{abstract}


\section{Introduction}

In one popular description of the Milky Way's interstellar medium, the
Sun lies within a cluster of small, cool clouds 
(Lallement \& Bertin 1992; 
Lallement et al. 1995; Linsky et al. 2000; Redfield \& Linsky
2000), which are themselves embedded within a large (approximately 70
pc in radius) bubble of highly ionized, rarefied gas (Cox \& Reynolds 1987;
Snowden et al. 1990; Warwick et al. 1993;
Frisch 1995; Snowden et al. 1998; Sfeir, et al. 1999).  
The bubble is called the Local Bubble or
the Local Hot Bubble and contains highly ionized gas, as evidenced by its
$\frac{1}{4}$ keV X-ray surface brightness and \oxysix\  absorption column
density (Sanders et al. 1977; Fried et al. 1980; McCammon et al. 1983;
Shelton \& Cox 1994; Oegerle et al. 2001).  Gas temperatures of
$10^{5.5}$ to $10^6$ K have been calculated from collisional
ionization equilibrium models of the observations.  However, it is
also possible that the gas has a lower temperature if it was heated by
an ancient explosion and has subsequently adiabatically expanded and
radiatively cooled (Breitschwerdt et al. 1996).

Beyond the Local Bubble lies the Lockman layer of neutral hydrogen
(Lockman et al. 1986), the Reynolds layer of ionized hydrogen
(Reynolds 1993), and the Galaxy's extended gaseous halo.  The Local
Bubble may be a bounded entity, though some recent evidence suggests a
connection with the halo (Welsh et al. 1999).  Like the Local Bubble,
the lower parts of the halo or the upper parts of the thick disk are
rich in highly-ionized gas.  \oxysix, \nitfive, and 
\carfour\ ions have been found
to extend to several kiloparsecs from the disk (Savage, Sembach, \&
Lu 1997; Savage et al. 2000; also Danly \& Kuntz 1993) and X-ray
shadowing observations show that some fraction of the
$\frac{1}{4}$ keV X-ray background
originates beyond the Lockman layer (Snowden
et al. 1991; Burrows \& Mendenhall 1991, Snowden et al. 2000).  
We do not know if these
ions reside in a hot plasma in collisional ionization equilibrium or if
they are the cooled tracers of previous heating events.
Such events may include halo
supernovae, shocks due to fast infalling clouds, and galactic
fountains.

Much of what is known about highly ionized gas in the Galactic
interstellar medium (ISM) has been derived from observations of soft
X-ray emission and absorption-line measurements of the \oxysix, \nitfive, and
\carfour\ ions.  Measurements of \oxysix\ emission can complement these data.
%Starting with Holberg's (1986) work with Voyager data,
%Radiation from \oxysix\ ions is the major form
%of cooling of $\sim10^5$ to $\sim10^6$ K gas and so 
%\oxysix\ 
%intensity measurements provide estimates of the cooling rate of the hot
%interstellar medium.
Furthermore,
the electron density and thermal pressure can be calculated from the
ratio of the doublet 
intensity to the absorption column density, while the plasma's ionization 
history can be inferred by comparing the \oxysix\ intensity and high 
resolution soft X-ray spectra.  
Many investigators have searched for
\oxysix\ emission from the diffuse interstellar medium, but the only previous
detection of Galactic \oxysix\ emission was made by the {\it{Hopkins
Ultraviolet Telescope}} ({\it{HUT}}).  In addition to observing
supernova remnants and the Loop I superbubble, {\it{HUT}} observed the
diffuse ISM toward $l = 218^\circ$, $b = 56^\circ$, detecting the \oxysix\
doublet in the wing of the terrestrial Ly$\beta$ emission with a
surface brightness of 19,000$\pm5000$ photons cm$^{-2}$ s$^{-1}$
sr$^{-1}$ (Dixon, Davidsen, \& Ferguson 1996).  Upper limits on the
\oxysix\ surface brightness have been established for other directions (see
Holberg 1986; Edelstein \& Bowyer 1993; Dixon, et al.
1996; Korpela, Bowyer, \& Edelstein 1998), with most finding
$\stackrel{<}{\sim} 7000$ photons cm$^{-2}$ s$^{-1}$ sr$^{-1}$.
Most recently, Edelstein et al. (1999) used
{\it{Espectrografo Ultravioleta extremo para la observacion de la 
Radiacion Difusa}}
({\it{EURD}}) data to find an upper limit of only 1200 photons
cm$^{-2}$ s$^{-1}$ sr$^{-1}$ with $90\%$ confidence for a swath along
the ecliptic, and Murthy et al. (2001) used low-resolution
{\it{Voyager}} data to find an upper limit of 2600 photons cm$^{-2}$
s$^{-1}$ sr$^{-1}$ with $90\%$ confidence toward $l = 117^\circ$, $b =
51^\circ$.

Here we report on the first observations of \oxysix\ ($\lambda\lambda
1032$ and 1038)
and \cartwo\ ($\lambda\lambda$ 1037) emission from the
diffuse ISM outside of supernova remnants or superbubbles
made with the {\it{Far Ultraviolet Spectroscopic Explorer}} 
({\it{FUSE}}).  
Dixon et al. (2001) have also used {\it{FUSE}} to observe 
diffuse \oxysix\ emission from
a sight line in the northern Galactic hemisphere.
{\it{FUSE}} has
sufficient spectral resolution to resolve easily the \oxysix\ emission lines
from \hone\ Lyman $\beta$ and \oxyone\ airglow lines several~\AA\ away
(see Sahnow et al. (2000) for a characterization of {\it{FUSE}}'s 
performance).  
These observations provide important constraints on models for the
Local Hot Bubble and Galactic thick disk and halo.  

In section 2, we discuss the properties of the observed sight line,
including the soft X-ray emission and dust opacity in this direction.
In section 3, we describe the {\it{FUSE}} observations and data
reduction techniques.  Our measurements of the intensity, velocity,
and width of the observed \oxysix\ and \cartwo\ lines are presented in 
section 4.  In section 5, we calculate the
%measured \oxysix\ 1032 and 1038~\AA\ intensities, measured
%ratio of intensities, 
optical depth, intrinsic intensities, column
density, electron density, thermal pressure, and emitting depth.
%This is done for three cases of interest.  In the first, all
%values are found from the observations, but the uncertainties are large.
%In the second, the 
%observed \oxysix\ is assumed to reside in the Local Bubble and
%in the third case, the observed \oxysix\ is assumed to reside
%partly in the Local Bubble and partly in the thick disk and lower halo.
The results are compared with information gained from other observations.
We conclude that
some of the observed photons probably originate in a 
thin region associated with the periphery of the Local Bubble,
while the majority of the observed photons probably originate in
the thick disk and lower halo.  The path length of the emitting
gas in the thick disk and lower halo is probably between 
ten and several hundred parsecs, and thus is very small when
compared with the \oxysix\ scale height found by
Savage {\it{et al}} (2000).
Section 6 contains a summary of our conclusions.
%We summarize our work in section 6.

\section{Observed Sight Line}

%Our pointing direction is $l = 315\fdg00$, $b = -41\fdg33$.
Our pointing direction is $l = 315.00^\circ$, $b = -41.33^\circ$
(RA = 22:44:30.0, Dec = -72:42:00.0).  The
observed intensity may include contributions from the Local Bubble and
material out through
the Galactic halo.  Other than the Local Bubble, there are no known
supernova remnants or superbubbles in this part of the sky.  The
pointing direction is near the edge of the Magellanic Stream (Wakker
\& van Woerden 1997), but no emission is observed at the Stream's
velocity.

Soft X-ray data provide some information on the conditions of the
hot gas along the observed line of sight, although
%Any inferences made, however, require the proviso that 
the plasma producing the soft X-rays is probably somewhat hotter or
more highly ionized than the plasma producing the \oxysix\ photons.
Within a $\sim5^\circ$ radius of the {\it{FUSE}} pointing,
%Within 70 square degrees of the {\it{FUSE}} pointing 
the average {\it{ROSAT}} $\frac{1}{4}$~keV surface brightness is 
$850 \times 10^{-6}$ counts
s$^{-1}$ arcmin$^{-2}$ (Kuntz 2000).  Figure~1 shows the Snowden et
al.  (1997) $\frac{1}{4}$~keV map of the sky; the surface brightness at the
position of the \fuse\ pointing is similar to the average observed for
latitudes $b < -30^\circ$.  The map also shows that the pointing
direction is not toward an unusual portion of the southern sky.  The
observed region lies between extended regions of above- and below-
average soft X-ray intensities, where the X-ray intensity gradient
appears to be associated with absorption by the Galaxy's \hone\ layer
(see Figure~1).

{\it{IRAS}} maps of 100 $\mu$m emission 
(Wheelock et al. 1994)
show that an irregular patch of
far-infrared (FIR) cirrus, about 30$^\prime$ in diameter, overlaps our
30'' $\times$ 30''
{\it{FUSE}} field of view.  Atomic hydrogen maps (21 cm) yield N$_{\rm{H
I}}\sim$ 4 $\times 10^{20}$ cm$^{-2}$, although, due to the coarse
binning %($1.0^\circ \times 10^\circ$ bins) 
(Dickey \& Lockman 1990; accessed through SkyView at 
http://skyview.gsfc.nasa.gov/skyview.html), 
significant variations are possible.  Based
on published photometry and spectral classification we have also
derived estimates of the visual extinction in this region.  Using a
search radius of 60\arcmin, centered on our pointing direction, we
find 8 stars with spectral classification in the Michigan
Spectroscopic Survey (Houk \& Cowley 1975) and B-V measurements from
Tycho photometry (Hog et al. 2000).  Their spectroscopic and parallax
information is listed in Table~1.
%Table~\ref{ebv}.
Using spectroscopic parallaxes we find weak evidence for an obscuring
``cloud'' at $\sim$200 pc causing a color excess of as much as 
${\rm E(B-V)} \sim0.07$ 
magnitude, with no evidence for further reddening out to
$\sim750$ pc.  This reddening is consistent with the column density
derived from the 21 cm data (Diplas \& Savage 1994), and we assign it
to the FIR (100~$\mu$m) cirrus patch.  

Far-ultraviolet emission
(e.g., from \oxysix\ 1032 and 1038~\AA) originating from beyond the cirrus
will be subject to scattering and absorption by the cirrus.
However,
because both the emission source and the cirrus are probably
extended, many of the 1032 and 1038~\AA\ photons scattered out of the
beam may be replaced by 1032 and 1038~\AA\ 
photons scattered into the beam.
%The fraction scattered into the beam is not known.  
%Neither is the effect of absorption alone.  
%Because we do not know the specifics of the absorption and scattering
%in this direction, we take as limits $0\%$ and $60\%$ loss.  The
%latter value is calculated from the Fitzpatrick (1999) extinction
%curve for point sources assuming $R_V = 3.1$ and ${\rm E(B-V)} 
%\sim0.07$.
Because we do not know the details of either the source geometry or the
FUV scattering characteristics in this direction, we can only estimate
the effective amount of extinction experienced by photons originating
from beyond the cirrus.  In one limiting case, 
there are losses due to both scattering and absorption,
so the average interstellar extinction curve can be applied to estimate
the extinction near 1032 and 1038~\AA.  
%Using 
%%Fitzpatrick and Massa's (1988) 
%Fitzpatrick's (1999)
%parameterization of the Fitzpatrick and Massa (1988) 
%extinction curve, R$_V$=3.1, and
%%E(B-V) $\leq 0.07$ from above, we find that A$_{1035} \leq 0.93$ mag.
%E(B-V) $\leq 0.07$ from above, we find that A$_{1035} \leq 0.99$ mag.
For E(B-V) $\leq$ 0.07 and R$_V$=3.1 (the standard value for 
the diffuse ISM), we find that the extinction curve and
parameterization presented in Fitzpatrick (1999) yields
A$_{1035} \leq 0.99$ mag.
Transforming this into an optical depth and applying the basic
radiative transfer equation, this yields an estimated loss of 
intensity of $\leq 60\%$.
In the other limiting case, all photons scattered out of
the beam are replaced by photons scattered into the beam.  In this
case, there are no net losses due to scattering but there are losses
due to absorption.  For the cases in which
scattering is inconsequential and absorption alone
is very small or the sight line traverses
a low density section of the cirrus,
we take as the extreme lower limit a 0$\%$ loss in intensity.
By combining these cases, we take as reasonable limits that
between $40\%$ and $100\%$ of the 1032 and 1038~\AA\ intensity
that enters the cirrus from beyond will emerge from the cloud and
arrive at the observer.

\section{Observations and Data Reduction}

The observations discussed here were acquired in 1999 September using
the low resolution (LWRS; 30\arcsec\ $\times$ 30\arcsec) aperture of
the {\it{FUSE}} instrument.  Summaries of the \fuse\ observatory and
on-orbit characteristics can be found in 
%Moos et al. (2000) and 
Sahnow et al. (2000).  The data sets used in this work are archived 
%at 
%the MultiMission Archive At the Space Telescope Science Institute
%(MAST) 
under the program identification I20509.  
This paper only presents data taken with
detector 1 because detector 2 was off during much of the observations.
After eliminating intervals when the spacecraft was in the South
Atlantic Anomaly and editing out event bursts (Sahnow et al. 2000),
the exposure durations with the 1A and 1B detector segments were 219
and 233 ksec, respectively.  Of these, 101 and 99 ksec, respectively,
were acquired while the satellite was in the night portion of its
orbit.  The differences between day and night data are discussed
below.

The data were processed using version 1.7 of the standard CALFUSE
pipeline (Sahnow et al. 2000; Oegerle, Murphy, \& Kriss 2000), with
four modifications:
%(1) the data were screened for event bursts, 
(1) the background noise originating in the detector was reduced by
excluding counts which produced pulseheights outside the expected
range for cosmic photons (the CALFUSE pipeline cutoffs were 
4 and 12 in the standard arbitrary units),
%(2) the CALFUSE astigmatism correction module was not used,
(2) the standard CALFUSE detector-background subtraction algorithm was not
used, allowing us to calculate the uncertainties in our measurements
more easily,
(3) we extracted the spectra from the spatially corrected
two-dimensional images of the detectors' signal in units of counts 
reported as an intermediate CALFUSE pipeline data product, and
in constructing the extraction windows, we used the 
the criterion that they must be extended far enough in the cross dispersion
direction to at least encompass the entirety of each 
terrestrial airglow diffuse emission signal,
(4) 
%effective area curves were used in the conversion from 
%the extracted spectra to units of ergs cm$^{-2}$ s$^{-1}$ sr$^{-1}$ 
%\AA$^{-1}$, but 
the LiF 1B point source effective area curve,
which corrects for the thin shadow of the repeller grid seen in point
source spectra over the $\sim1155$ to $\sim1170$~\AA\ range (also
known as the ``worm''), was replaced with an effective area curve that
does not include this correction.
%was replaced with a curve which did not correct for the ``worm'' because
%the ``worm'' does not affect the diffuse source spectra nearly as much 
%as it affects point source spectra. 
The relative wavelength scale of the resulting spectrum is accurate to
five or six detector pixels, or about 0.035~\AA\ in the LiF 1A spectrum.
The absolute wavelength scale is less well known, and we discuss the
determination of a wavelength (velocity) zero-point below.  The
intensity calibration was achieved using the best instrumental
affective area estimates as of June 2000 (Sahnow et al. 2000).  The
uncertainties in the solid angle of the aperture and 
effective area calibration are the largest sources of
systematic error.  The LWRS's solid angle is thought to be accurately known to
about $10\%$, while
the effective area calibration is thought to be
accurate to $\sim10\%$ for LiF 1A, SiC 1A, and SiC 1B, and $\sim20\%$
for the LiF 1B outside of the region of the ``worm.''

Figure~2 presents the
%Figure~\ref{totalspectrum} presents the 
spectra extracted from the detector 1 data in counts versus wavelength
(i.e., no intensity calibration has been performed).  LiF and SiC
refer to the lithium fluoride and silicon carbide mirror and grating
coatings used on the respective channels.  The strongest emission
lines seen in Figure~2 come from the Earth's atmosphere.  Across
most of the bandpass, we see only noise due to the detector
background, light scattered within the spectrograph, and cosmic
background radiation.
%The intensity gradients at short wavelengths in the LiF 1A spectrum
%and long wavelengths in the LiF 1B spectrum
%are due to gradients in the scattered light levels and
%LiF mirrors' effective areas.
The small elevations in the intensity between $\sim1045$ and
$\sim1055$~\AA\ in the LiF 1A spectrum and between $\sim1025$ and
$\sim1035$~\AA\ in the SiC 1A spectrum are due to light scattered
within the spectrograph during the day portion of the orbit.  This
scattered light makes it untenable to use the day time SiC 1A data for
analysis of the \oxysix\ emission.  The night-only portion of the data
provides less than half of the total exposure time.  For this reason
and because the SiC 1A channel has a smaller effective area at 
1032~\AA\ than does the LiF 1A channel, the SiC 1A data are not used in 
the \oxysix\ analysis.  However, the night SiC 1B data are used to search
for \sulsix\ 933 and 944~\AA\ and \carthree\ 977~\AA\ emission since the LiF
data do not cover these wavelengths.
As can 
be seen in Figure 2, the LiF 1A data show a peak in the vicinity of 1032
A that is potentially an \oxysix\ line, and another near 1037-1038 A that
potentially contains a \cartwo\ line and the other \oxysix\ line.

%Henceforth, only the LiF 1A data will be used for the \oxysix\ analysis.  
%However, being the only data on possible S VI emission at 933 and 945
%\AA\ and \carthree\ emission at 977~\AA, the night-time SiC 1A observation
%is used in the search for these emission lines.

%The light scattered onto the $\sim$1045 to $\sim$1055~\AA\ portion of the
%LiF 1A spectrum and $\sim$1025 and $\sim$1035~\AA\ portion of the SiC 1A 
%has been observed in other spectra as well.  Combined data sets show that
%the scattered light fell on the detector in a weak pattern of
%nearly vertical stripes.
%Figure~\ref{ovidayvsnight} shows the $\sim1020$ to $\sim1040$~\AA\ region
%of the LiF 1A intensity spectra.
We considered the possibility that some of the peaks observed in the 1030
to 1040~\AA\ region of the LiF 1A spectrum could be due to light
scattered within the spectrograph.
%or telescope 
Such scattering would be expected to distribute some light
in the cross dispersion direction,
%beyond the focal plane assembly 
as with the $\sim1045$ to $\sim1055$~\AA\ contamination, 
which would be visible outside of the spectral extraction
region.  However, there is
no excess of counts 
%above or below 
%outside of the extraction window 
perpendicular to the dispersion direction
in this region of the LiF 1A spectrum.  Thus, we rule out this type of
scattering.
%Thus the region is probably not contaminated
%by light scattered within the instrument beyond the spectrograph.
%Furthermore, the strongest sources of potentially scattered light are
%the Sun and the atmosphere on the side of the Earth facing the Sun.
%Observations of such light would be orders of magnitude stronger during
%satellite-day than during satellite-night.  However,  the observed
%intensity in the 1030 to 1040~\AA\ region

We excluded the possibility that the observed 1032, 1037, and 
1038~\AA\ emission lines could have come from
the Earth's atmosphere because of their continued strong emission in
the satellite-night portion of the observations.  Atmospheric airglow
emission (such as the \hone\ Lyman-$\beta$ and \oxyone\ emission lines at rest
wavelengths of 1025.7, 1027.4, 1028.2, 1039.2, 1040.0, 1040.9~\AA)
are roughly an order of magnitude stronger during the
satellite-day portions of the observation than during the
satellite-night portions.  
In contrast,
%As Figure~3 shows,
%As Figure~\ref{ovidayvsnight} shows,
the \oxysix\ and \cartwo\ emission features in our data
have constant intensity levels
between the satellite-night and satellite-day portions of the
observation.  Thus, the emission features at 1032, 1037, and 
1038~\AA\ are not due to terrestrial airglow (see Figure~3).
%Thus, the observed features cannot be attributed to atmospheric
%airglow.
%while
%the strengths of the nearby \hone\ Lyman $\beta$ and \oxyone\ emission lines at 
%1025.7, 1027.4, 1028.2, 1039.2, 1040.0, 1040.9~\AA\ .

We examined the possibility that scattered solar emission lines could
have produced the features observed in the LiF 1A data.  The Sun's
ultraviolet spectrum has been recorded by the {\it{Solar Ultraviolet
Measurements of Emitted Radiation}} ({\it{SUMER}}) instrument aboard the
{\it{Solar and Heliospheric Observatory}} ({\it{SOHO}}).  The two
strongest emission lines common to the SUMER bandpass
and the combined {\it{FUSE}} LiF 1A and LiF 1B
bandpass are the \oxysix\ 1031.9~\AA\ line and the 1175.7~\AA\ member of
the \carthree\ triplet (Curdt et al. 1997).  In the solar spectrum,
%the \oxysix\ line is 1.17 times stronger than the \carthree\ 1175.7 line.  
the \carthree\ 1175.7~\AA\ line is 0.86 times as bright as the \oxysix\ 
1032~\AA\ line.  Thus, if the 1032~\AA\ photons observed by {\it{FUSE}}
came from the Sun, then {\it{FUSE}} should have also observed 
1175.7~\AA\ emission with 86$\%$ the intensity level of the 1032~\AA\
emission.  But, the {\it{FUSE}} spectrum contains no emission features
within at least 1~\AA\ of 1175.74~\AA.  For $\lambda = 1175.74 \pm
0.2$~\AA, the intensity recorded by {\it{FUSE}}, and measured with the
techniques described below, is 360 $\pm$ 400(random) photons s$^{-1}$
cm$^{-2}$ sr$^{-1}$.
%This is far less than that measured
%for the 1032~\AA\ emission line: $2900 \pm 270$
%photons s$^{-1}$ cm$^{-2}$ sr$^{-1}$.
This provides upper limits to the possible solar contamination of the
LiF 1A spectrum at 1032 and 1038~\AA\ of $420 \pm 470$(random) and 
$210 \pm 230$(random) photons s$^{-1}$ cm$^{-2}$ sr$^{-1}$, respectively.
%Thus, the
%observed \oxysix\ emission is not due to solar contamination.

The LiF mirrors are physically situated on the shadowed side of the
satellite, while the SiC mirrors are on the sun-illuminated side.
This motivated us to search for possible solar contamination in
the SiC data.  The \carthree\ 977~\AA\ emission line observed in the solar
spectrum was also found in the day time SiC 1B spectrum but was not
found in the night time SiC 1B spectrum.  The observed intensity is
more than can be explained by terrestrial airglow.  Thus, we conclude
that the SiC 1 data taken during the day portion of the satellite's
orbit may contain solar contamination, but the data taken at night
contains no significant contamination.

\section{Spectral Analysis}

\subsection{\oxysix\ Emission}

%\noindent{\it{\oxysix\ Emission:}}

%Table~\ref{ovimeasurements} presents

Table~2 presents the measured intensities and uncertainties, $\sigma$,
the central velocities, the deduced intrinsic full width at half
maximum values, and the implied kinetic temperatures for the two \oxysix\
features.  
The values given in Table~2 were determined as follows.  The
background continuum was found by fitting a polynomial to the spectrum
outside of the \hone\ Ly$\beta$ airglow, \oxyone\ airglow, and next four
strongest emission features (the 1032~\AA\ 
\oxysix\ emission line, the 1038~\AA\ \oxysix\ emission line, the 
1037~\AA\ \cartwo\ emission line, and a spurious $\sim$1031~\AA\ feature).
%Spectra for other wavelength regions were measured
%using a similar procedure.
The continuum was subtracted from the spectrum. The intensities of the
\oxysix\ emission features were measured by summing the signal over each
line.
%The measurement of the standard deviation
%was obtained from the counts spectrum and converted to photon units.
%The standard deviation was measured from the counts spectrum and 
%converted to photon units.
%We calculated each feature's standard deviation from the pixel to pixel
%root-mean-square measured between 1029 and 1038~\AA\
%scaled by the square root of the ratio of the number of pixels
%per emission feature to the number of pixels between 1029 and 1038~\AA.
The statistical (random) $1\sigma$ uncertainty in the number of photons 
in each feature was
calculated as the square root of the sum of the signal and background.
%as the square root of the product of the number of pixels
%in the feature and the mean variance per pixel, relative to the
%background continuum fit over the region 1029 to 1038~\AA.  
%Although the spectral regions with \oxysix\ and \carthree\ 
%emissions were included
%in this, they did not significantly increase the variance.
%The uncertainty was then converted to photon units.  
The $1\sigma$ systematic uncertainty was calculated by adding in
quadrature the $10\%$ uncertainties in the aperture solid angle and flux
calibration.
The wavelength scale
was corrected using an interpolation pinned to the rest wavelengths of
the four strongest nearby airglow
\hone\ and \oxyone\ 
%H Ly $\beta$ and \oxyone\ 
emission lines (Morton 1991, 2001) and the CALFUSE wavelengths at
which these features were observed.  The corrections applied to the
1032 and 1038~\AA\ emission features were $-0.24$ and $-0.23$~\AA,
respectively.  We then calculated the velocities relative to the
geocenter and the Local Standard of Rest (LSR).  The wavelength scale is
probably accurate to within several detector pixels, or about 
10 km s$^{-1}$, and the centroid determination is probably accurate to within
about 10 km s$^{-1}$ as well.
%Gaussian curves were fit to the observed
%emission features.
%The purpose of this procedure was to estimate the positions
%of the features' centers and the features' apparent widths. 
The line centroids 
%and apparent widths 
were found by fitting Gaussian functions to the profiles.
The observed widths are greater than the widths of the underlying
cosmic spectrum because the {\it{FUSE}} instrument convolves the
intrinsic spectrum with the instrumental point spread function and
the large slit function.  
%The instrumental spread function can be
%approximated as a $0.365$~\AA\ (106 km s$^{-1}$) wide top hat function
%(e.g., see the \oxyone\ feature shown in Figure~4).  
In an analysis of unblended \oxyone\ and \arone\ airglow lines in this and 
other data sets, we found that 
the instrumental effects are almost completely responsible for the 
the shapes and widths of the unblended airglow features
(e.g., see the \oxyone\ feature shown in Figure~4).
Thus, we take the nearby 1039~\AA\ \oxyone\ airglow feature's profile as the
net instrumental function.
We convolved this function with Gaussian functions of various widths,
then compared the results with the observed \oxysix\ profiles in order to
determine the intrinsic widths of the \oxysix\ emission lines.  
By equating the
line dispersion with a thermal width we were able to calculate an upper
limit on the kinetic temperature of the emitting plasma.

The measured intensities of the \oxysix\ 1032 and 1038~\AA\ emission 
features are 2930$\pm$290(random)$\pm$410(systematic) and 
1790$\pm$260(random)$\pm$250(systematic) photons cm$^{-2}$ s$^{-1}$ 
sr$^{-1}$, corresponding to detections 
at $10\sigma$ and $7\sigma$ levels of significance with respect to
the random errors.  Henceforth, the $\sigma$ associated with the
random errors will be denoted by ``(r)'' and the $\sigma$ associated
with the systematic errors will be denoted with by ``(s)''.
%The observed intensity is a fourth
%as strong as the HUT measurement of \oxysix\ emission from the ISM
%in the northern Galactic polar region outside of known supernova
%remnants and superbubbles.  
The velocity profiles of these lines are shown in Figure~4.
For comparison, the (unresolved) \oxyone\ airglow profile is also plotted.
The 1032 and 1038~\AA\ features have similar central velocities,
observed and intrinsic widths, and the observed velocity profiles have
roughly similar shapes.  The agreement corroborates our conclusion
that the 1032 and 1038~\AA\ features are produced by a single species.

The observed ratio of the intensities in the 1032~\AA\ and 1038~\AA\
lines (calculated from unrounded values; the values
in Table~2 were rounded to 3 significant digits) is 1.64$\pm$0.29(r).
The systematic uncertainties affect the 1032 and 1038~\AA\ emission 
line measurements in the same manner and thus the ratio of the 
1032 and 1038~\AA\ intensities is not affected by the systematic 
uncertainties.
%The ratio and $\sigma$ were calculated from the unrounded values.
%If the emitting plasma is optically thin, then the ratio should equal
%that of the statistical weights (2.0), and if the emitting plasma is
%optically thick, then self absorption should lower the ratio to 1.0.
%The observed ratio lies midway between and within $\sim2\sigma$ of the
%optically thin and optically thick ratios.  
In the limit of the optically thin case, as the optical depth 
approaches zero, the
ratio should approach that of the statistical weights, 2.
In the optically thick case, self absorption
may change the ratio. The resulting ratio depends on
the optical depth and the geometry of the \oxysix-rich material.
The ratio that we have observed lies 1.2 $\sigma$ 
below the theoretical optically thin value.
The upper limits on the
kinetic temperature surpass the temperature at which \oxysix\ is most
abundant in collisional ionization equilibrium plasma, but the breadths of
the \oxysix\ profiles could in large part be due to non-thermal
processes such as turbulence or dispersion in bulk motions.

%Possible causes for the breadth of the emission features are 
%high temperatures and large velocity gradients in the emitting 
%plasma.

%\vspace{0.5cm}
\subsection{\cartwo\ Emission}
%\noindent{\it{\cartwo\ Emission:}}

We have also observed emission from the 3s $^2S_{1/2}$ to 2p $^2P_{3/2}$
transition of \cartwo\ ($\lambda = 1037.02$~\AA).  The emission appears to
be centered at a corrected LSR wavelength of 1037.16~\AA.  The emission
probably arises from the $^2S_{1/2}$ to $^2P_{3/2}$ transition rather
than the $^2S_{1/2}$ to $^2P_{1/2}$ transition to ground state \cartwo\
($\lambda = 1036.34$~\AA).  This is probably because (1) the apparent
wavelength is much closer to
%the rest wavelength of \cartwo\ than the rest wavelength of \cartwo, 
$1037.02$~\AA\ than to $1036.34$~\AA,  (2) the A value for the
$^2S_{1/2}$ to $^2P_{3/2}$ transition is more than twice that of the
$^2S_{1/2}$ to $^2P_{1/2}$ transition, and (3) the ISM's plentiful
population of ground state \cartwo\ should readily absorb 
1036.34~\AA\ photons.

%Table~\ref{ciimeasurements} 
Table~3 presents the observed intensity, $\sigma$, central LSR
velocity, and width of the Gaussian fit to the 1037~\AA\ feature.  The
emission feature is redshifted relative to rest, but to a slightly
lesser degree than the \oxysix\ lines.  The \cartwo\ feature is also
significantly narrower than the \oxysix\ features, as might be expected
from gas in a lower ionization stage.  The observed FWHM is similar to
the instrumental FWHM for diffuse emission that fills the large
aperture, i.e., the \cartwo\ emission line is poorly resolved.

%\vspace{0.5cm}
%\noindent{\it{Confirming the \oxysix\ and \cartwo\ Line Identifications:}}
\subsection{Search for H$_2$ Fluorescence between 1031 and 1038~\AA:}

Hydrogen molecules in high rotational states have several possible
transitions that would produce 1032 to 1038~\AA\ photons.  
We have applied two tests to
confirm that the emission lines observed by {\it{FUSE}} are
indeed those of
the \oxysix\ doublet and \cartwo\ ($^2S_{1/2}$ to $^2P_{3/2}$) rather
than fluorescing H$_2$.
%To confirm the identities of the 1032, 1038, and the 1037~\AA\ emission peaks
%as those of the \oxysix\ doublet and excited \cartwo\, we tested the possibility
%that they could be from fluorescing hydrogen molecules.
In the first test, we set limits on the strength of the H$_2$ lines
in the vicinity of 1032~\AA\ and 1038~\AA\ by finding other lines
in the {\it{FUSE}} bandpass that would be excited by the same
mechanism and using the {\it{FUSE}} spectrum to measure their strengths.
Morton \& Dinerstein (1976), Sternberg (1989), and Abgrall et
al. (1993a,b) list the following H$_2$ emission lines:
%P(3) Lyman (6,0) transition at 1031.19~\AA, 
%P(2) Werner (1,1) transition at 1031.36~\AA, 
%Q(3) Werner (1,1) transition at 1031.87~\AA, 
%R(4) Lyman (6,0) transition at 1032.35~\AA,
%%the P(3) Werner (1,1) transition at 1033.66~\AA, 
%R(1) Lyman (5,0) transition at 1037.15~\AA,  
%and P(1) Lyman (5,0) transition at 1038.16~\AA.
Lyman (6,0) P(3) transition at 1031.19~\AA, 
Werner (1,1) P(2) transition at 1031.36~\AA, 
Werner (1,1) Q(3) transition at 1031.87~\AA, 
Lyman (6,0) R(4) transition at 1032.35~\AA,
%the P(3) Werner (1,1) transition at 1033.66~\AA, 
Lyman (5,0) R(1) transition at 1037.15~\AA, and Lyman (5,0) P(1)
transition at 1038.16~\AA.  In order for an H$_2$ molecule to make the
Lyman (6,0) P(3) transition, the ambient radiation field must have
first pumped the molecule to the J=2 rotational state, v$^\prime$ = 6
vibrational state, and B$^1\sum_u$ electronic excitation state.  The
strength of the Lyman (6,3) P(3) (1172.38~\AA) transition arising from
the same upper state should be 58\% as strong as the Lyman (6,0) P(3)
transition at 1031.19~\AA (Abgrall et al. 1993a).
%From that state, the molecule is 
%35\% as likely to de-excite
%via the P(3) Lyman (6,1) transition and emit a 1077.01~\AA photon.  
%30\% as likely to de-excite
%via the R(1) Lyman (6,1) transition and emit a 1070.58~\AA photon.  
%58\% as likely to de-excite
%via the P(3) Lyman (6,3) transition and emit an 1172.38~\AA\ photon.  
%and 44\% as likely to de-excite
%via the R(1) Lyman (6,3) transition and emit an 1165.54~\AA\ photon.  
We have searched the wavelength-corrected {\it{FUSE}} spectrum for
emission at this wavelength having the width of the nearest airglow
line, finding an integrated intensity of -1300 $\pm$ 620 photons
cm$^{-2}$ s$^{-1}$ sr$^{-1}$.  Thus, fluorescing H$_2$ molecules
probably are not contributing 1031.19~\AA\ photons.  We have applied
this method to each of the H$_2$ transitions above.  In every case,
the signal is less than 2$\sigma$.  We used the Sternberg (1989) model
for our second test.  In Sternberg's spectrum of H$_2$ lines emitted
by cool clouds bathed in ultraviolet light, the Q(1) Werner
(0,0) line at $\sim 1009.77$~\AA\ and the P(3) Werner (0,1) line at
$\sim 1058.82$~\AA\ should be at least 138\% and 152\% as bright as
any of the H$_2$ emission lines between 1030 and 1040~\AA.  We have
measured the intensities of these comparison lines, finding integrated
intensities of 39 $\pm 110$ and 72 $\pm 170$ photons cm$^{-2}$
s$^{-1}$ sr$^{-1}$, respectively.  Thus, H$_2$ fluorescent emission
is unlikely to be responsible for the intensities observed near 1032, 1037,
and 1038~\AA.

%\vspace{0.5cm}
\subsection{Upper Limits on Other Emission Lines:}
%\noindent{\it{Upper Limits on Other Emission Lines:}}

We searched the LiF 1A and 1B 
spectra for interstellar emission from \carone\ at
1122.260~\AA, Si~{\footnotesize{II}} at 1023.700~\AA,
%P~II at 1156.970~\AA, 
S~{\footnotesize{III}} at 1015.502~\AA,
S~{\footnotesize{IV}} at 1062.664~\AA, and
Fe~{\footnotesize{III}} at 1122.524~\AA\ 
and we searched the night time SiC 1B spectra for interstellar emission
from S~{\footnotesize{VI}} at 933.378 and 944.523~\AA\ and 
\carthree\ at 977.020~\AA.  No
evidence for emission from any of these lines was found.  The
resulting 3$\sigma$ upper limits are listed in Table~4.
%Table~\ref{otherspecies}.  

%\pagebreak

%\section{Discussion}
\section{Physical Characteristics}

%We start with a necessarily vague understanding of the
%emitting gas.  Its 
%%size and shape, 
%size, shape, \oxysix\ column density, electron density, and thermal pressure
%are not known.  
%From the observed intensities of the 1032 and 1038~\AA\ lines,
%estimates for extinction by dust, and ... the ratio of observed
%intensities ...
%
%The 1032 and 1038~\AA\ photons produced
%by the \oxyfive\ ions in the gas may scatter off of or
%be absorbed by dust within the region or between the emitting
%region and the observing instrument.
%The 1032 and 1038~\AA\ photons 
%may also scatter off of 
%other \oxyfive\ ions within the region.  
%If photons scatter off of \oxyfive\ ions in the region, 
%then the intensity along a given line of sight
%may be increased or decreased depending on whether more
%photons scattered into the line of sight than scattered out.
%The 1032~\AA\ photons are more likely to be scattered than
%the 1038~\AA\ photons, so scattering can 
%affect the ratio of the intensities of these photons 
%along a given line of sight.
%However, scattering off of \oxyfive\ ions does not
%affect the photon flux integrated over the entire surface area.
%
%The size and shape of the emitting region are not known,
%{\it{a priori}}.  

In this section, we derive equations that relate the observed
intensities of the \oxysix\ emission lines, the optical depths of the
emitting regions in these lines, the intrinsic intensities of
their emission, and the column densities of the lines. Additional
equations relating the intrinsic emission intensity and the column
densities to the electron density, thermal pressure and physical extent
of the emitting region or regions are also presented.
We then calculate the physical characteristics for the following cases of
interest.
In cases $\#1a$ and $\#1b$, the above quantities are
calculated from the observed line intensities and their ratio.
In case $\#1a$, we assume that there is no extinction by the possible
cirrus material.  In case $\#1b$, we assume that the cirrus reduces the
1032 and 1038~\AA\ intensities by $60\%$.
The uncertainties in the values calculated for cases $\#1a$ and $\#1b$ are
very large, primarily owing to the large uncertainties in the
line ratio.  Thus we also consider cases $\#2$, $\#3a$, and $\#3b$,
which draw the \oxysix\ column density from assumptions about the
physical scenario rather than the line ratio.
In case $\#2$, the sole source of the observed emission is assumed
to be the Local Bubble and therefore an estimate appropriate for the Local
Bubble is taken for the column density.  
%The other
%characteristics are calculated from the observed doublet intensity
%and assumed column density.
No cirrus extinction is assumed for case $\#2$ because the
cirrus material is thought to lie beyond the Local Bubble.
In  cases $\#3a$ and $\#3b$, the observed emission is assumed to originate 
partly in the Local Bubble and partly in the Galactic thick disk and
lower halo.
Therefore an 
appropriate estimate is taken for the column density.
%The other
%characteristics are calculated from the observed doublet intensity
%and assumed column density.
In case $\#3a$, we assume that there is no extinction by the possible
cirrus material.  In case $\#3b$, we assume that the cirrus reduces
the 1032 and 1038~\AA\ intensities by $60\%$.  Cases $\#2$ and $\#3$ 
are probably the most plausible.

After presenting the results for each of these cases, we 
discuss five significant assumptions 
made in this analysis (especially the assumptions of negligible
scattering of 1032 and 1038~\AA\ photons into the beam), 
consider their validity, and discuss the 
sensitivity of our results to these assumptions.
Finally, we use the various derived physical characteristics to constrain
the possible sources for the observed emission.
%In the former case,
%the resulting intrinsic intensity and thermal pressure are much larger 
%than expected for the Local Bubble, while in the latter case,
%the results are acceptable for the 
%Galactic disk and halo case.  This leads us to tentatively assign the
%majority of the emission to the Galactic thick disk and halo, although 
%this assignation is weakened by the puzzling findings that the 
%observed velocities cannot be easily explained by either choice.

%\subsection{Physical Conditions}
\subsection{Derivations}

Our first goal is to determine how the observed intensities in
the 1032 and 1038~\AA\ lines are related to the intrinsic intensities.  
By intrinsic intensity, we mean the intensity of photons produced
in the plasma.  This intensity would have been
observed if resonant scattering and extinction by dust had not
changed the number of photons heading toward our detector.

By using a self absorption analysis, we are able to link the observed and
intrinsic intensities to one another via the optical depth term,
and find a ready explanation for the difference between the 
observed line ratio
and the optically thin value.  Note, however, that the
difference between the observed line ratio and the optically thin
value is only marginally statistically significant.
In the self absorption scenario, photons emitted by the \oxysix\
ions may be scattered out of the beam by interactions with
other \oxysix\ ions within the emitting region
and residing along the line of sight, while only
negligible numbers of photons produced by other
\oxysix\ ions in the emitting region are scattered into the beam.
In section 5.3, we question
this assumption and discuss how the calculated quantities would be affected
if non-negligible numbers of 1032 and 1038~\AA\ photons
%produced by other \oxysix\ ions in the diffuse ISM
were scattered into the beam.

The standard radiative transfer equations governing emission produced
within a slab of material for the case of negligible scattering
into the beam and negligible extinction by dust are
Spitzer's (1978)
equations (3-1), (3-2), and (3-3) for the case where no externally
produced radiation 
is incident upon
the slab and $j_\nu/\kappa_\nu$ is constant with
respect to $\tau_\nu'$.
These equations
yield the observed intensity per frequency interval for a
given emission line:
\begin{equation}
I_{o\nu} = \int_0^{\tau_\nu} \frac{j_\nu}{\kappa_\nu} e^{-\tau_{\nu}'} 
d\tau_{\nu}' =
\frac{j_\nu}{\kappa_\nu} (1 - e^{-\tau_\nu})
\end{equation}
and the intrinsic intensity per frequency interval:
\begin{equation}
I_{i \nu} = \int_0^{\tau_\nu} \frac{j_\nu}{\kappa_\nu} d\tau'_\nu = 
\frac{j_\nu}{\kappa_\nu} \tau_\nu,
\end{equation}
%See Spitzer's equations (3-1) and (3-2) for the intrinsic intensity.
where $\nu$ is the frequency, 
$I_\nu$ is the specific intensity per unit frequency,
%in units of ergs cm$^{-2}$ s$^{-1}$ sr$^{-1}$ Hz$^{-1}$, 
$j_\nu$ is the emissivity coefficient,
%in units of ergs cm$^{-3}$ s$^{-1}$ sr$^{-1}$ Hz$^{-1}$, 
$k_\nu$ is the absorption coefficient,
%in units of cm$^{-1}$,
%$s$ is the distance variable,
$\tau'_\nu$ is the optical
depth variable, and $\tau_\nu$ is the total optical depth through the
emitting region.
Following Spitzer, in our radiative transfer derivations
$I_\nu$ is
in units of ergs cm$^{-2}$ s$^{-1}$ sr$^{-1}$ Hz$^{-1}$, 
$j_\nu$ is in units of ergs cm$^{-3}$ s$^{-1}$ sr$^{-1}$ Hz$^{-1}$, 
and 
$k_\nu$ is in units of cm$^{-1}$.  
However, when presenting the results, we give the
specific 
intensity in units of photons cm$^{-2}$ s$^{-1}$ sr$^{-1}$ Hz$^{-1}$.
% and in equation (10) in order
%to be consistent with
%Shull \& Slavin's(1994) original expression.

We combine Spitzer's equations (3-16), (3-23), and (3-25) to yield
an expression for $\tau_\nu$.
The gas is assumed to be far from equivalent thermodynamic equilibrium
though it may be in collisional ionizational equilibrium.  
Thus Spitzer's $\frac{b_k}{b_j}$ can be taken as zero, because
the populations of the upper levels are negligible.  
Thus,
\begin{equation}
%\tau_\nu = N_j s <\phi(\Delta\nu)> = N_j \frac{\pi e^2 f_{jk}}{m_e c} 
%(1 - \frac{b_k}{b_j}e^{-h\nu/{\rm{k}}T}) <\phi(\Delta\nu)>,
\tau_\nu 
%= N_j s <\phi(\Delta\nu)> 
= N_j \frac{\pi e^2 f_{jk}}{m_e c} 
<\phi(\Delta\nu)>,
\end{equation}
where 
$N_j$ is the column density of ions in the $j$ state,
%$s$ is the absorption cross section per particle,
$\Delta \nu = \nu - \nu_{o}$,  
%{\bf{(note, this disagrees with $\Delta \nu$ notation used later!)}},
%I commented out the old Delta nu notation.
$\nu_{o}$ is the frequency at line center,
%$\phi(\Delta \nu)$ is the normalized distribution with respect to frequency,
$<\phi(\Delta \nu)>$ is the line profile function (the
path length-averaged normalized
distribution with respect to frequency of the cross section per
\oxysix\ ion),
$e$ is the charge of the electron,
$f_{jk}$ is the oscillator strength for the chosen transition, 
$m_e$ is the mass of the electron, 
and $c$ is the speed of light.
%$b_j$ is the ratio of the actual density of ions in the $j$ state to the
%``equivalent thermodynamic equilibrium'' density of such particles,
%$b_k$ is the equivalent ratio for particles in the $k$ state, 
%$h$ is Planck's constant,
%${\rm{k}}$ is Boltzman's constant,
%and $T$ is the temperature.
%Thus, $\frac{m_e c}{\pi e^2} = 37.68$.
From Spitzer's equation (3-25), 
$\frac{\pi e^2}{m_e c} = 2.654 \times 10^{-2}$,
where the units should be $\frac{\rm{cm}^2}{\rm{s}}$.
If the particle velocity distribution is Maxwellian, then 
$<\phi(\Delta \nu)> = \frac{\lambda}{b \sqrt{\pi}} 
e^{-(\frac{\lambda \Delta \nu}{b})^2}$.
Here $b$ is the velocity-spread parameter, 
$\Gamma_v$ is the FWHM velocity, and $b = 0.6 \Gamma_v$.
If we define $\tau_o$ as the optical depth at line center, then 
\begin{equation}
%\tau_\nu = N_j \frac{\sqrt{\pi} e^2 f_{jk} \lambda}{m_e c b} 
%(1 - \frac{b_k}{b_j} e^{-h\nu/{\rm{k}}T}) 
%e^{-(\frac{\lambda \Delta \nu}{b})^2} 
%= \tau_o e^{-(\frac{\lambda \Delta \nu}{b})^2},
\tau_\nu = N_j \frac{\sqrt{\pi} e^2 f_{jk} \lambda}{m_e c b} 
e^{-(\frac{\lambda \Delta \nu}{b})^2} 
= \tau_o e^{-(\frac{\lambda \Delta \nu}{b})^2}.
\end{equation}

We now integrate equations (1) and (2) with respect to $\nu$ in order
to determine $I_o$ and $I_i$.
Assuming that $j_\nu/\kappa_\nu$ is constant 
over the frequency interval where $1 - e^{-\tau_\nu}$ differs from $0$,
\begin{equation}
I_o = \frac{j_{\nu_o}}{\kappa_{\nu_o}} \int_0^\infty 
(1 - e^{-\tau_\nu}) d\nu
%= \int_0^\infty \frac{j_\nu}{\kappa_\nu} d\nu \times \int_0^\infty 
%(1 - e^{-N_j \frac{\sqrt{\pi} e^2 f_{jk} \lambda}{m_e c b} 
%(1 - \frac{b_k}{b_j} e^{-h\nu/{\rm{k}}T}) e^{-(\frac{\lambda 
%\Delta \nu}{b})^2}
%}) d\nu 
%= 2 \int_0^\infty \frac{j_\nu}{\kappa_\nu} d\nu \times \int_0^\infty 
%(1 - e^{-N_j \frac{\sqrt{\pi} e^2 f_{jk} \lambda}{m_e c b} 
%(1 - \frac{b_k}{b_j} e^{-h\nu/{\rm{k}}T}) e^{-(\frac{\lambda 
%\Delta \nu}{b})^2}
%}) d(\Delta\nu)
= \frac{2 j_{\nu_o}}{\kappa_{\nu_o}} \int_0^\infty 
(1 - e^{-\tau_o e^{-(\frac{\lambda \Delta \nu}{b})^2}
}) d(\Delta\nu).
\end{equation}
Equation (4) of Jenkins (1996) provides
a solution to the integral in our equation (5) for 
$\tau_o \stackrel{<}{\sim} 15$:
$F(\tau_o) = \int_0^\infty [1 - e^{(-\tau_o e^{-x^2})}] dx 
= \frac{\sqrt{\pi}}{2} \sum_{n=1}^{\infty} \frac{(-1)^{n-1} \tau_o^n}
{n! \sqrt{n}}$.
For 
%$\tau_o = N_j \frac{\sqrt{\pi} e^2 f_{jk} \lambda}{m_e c b} 
%(1 - \frac{b_k}{b_j} e^{-h\nu/{\rm{k}}T})$, and 
$x = \frac{\lambda \Delta \nu}{b}$, we find
\begin{equation}
I_o = \frac{\sqrt{\pi} b}{\lambda} \frac{j_{\nu_o}}{\kappa_{\nu_o}} 
 \sum_{n=1}^{\infty} \frac{(-1)^{n-1} \tau_o^n}{n! \sqrt{n}} =
\frac{2 b}{\lambda} \frac{j_{\nu_o}}{\kappa_{\nu_o}} F(\tau_o).
\end{equation}
%$I_o = A F(t) = A \int_0^\infty [1 - e^{(-t e^{-x^2})}] dx$,
%where $A = 2 \frac{j_\nu}{\kappa_\nu} \frac{b}{\lambda}$
%and $t = N_j \frac{\sqrt{\pi} e^2 f_{jk} \lambda}{m_e c b} 
%(1 - \frac{b_k}{b_j} e^{-h\nu/{\rm{k}}T})$.
%From Jenkins's equation (4),
%$I_o = A \frac{\sqrt{\pi}}{2} \sum_{n=1}^{\infty} 
%\frac{(-1)^{n-1} t^n}{n! \sqrt{n}}$
%for values of $t < 15$ 
%Integrating equation {\bf{(fill in)}} yields:
If there is dust between the emitting plasma and the observer,
then equation (6) must be modified to account for the 
losses due to extinction.  
Equation (6) becomes:
\begin{equation}
I_o = \frac{a 2 b}{\lambda} \frac{j_{\nu_o}}{\kappa_{\nu_o}}
% = \frac{a \sqrt{\pi} b}{\lambda} \frac{j_{\nu_o}}{\kappa_{\nu_o}}
% \sum_{n=1}^{\infty} \frac{(-1)^{n-1} \tau_o^n}{n! \sqrt{n}},
F(\tau_o),
\end{equation}
where $a$ is the ratio of the intensity
that emerges at the near side of the dust to the intensity that impinges
upon the far side of the dust. As discussed in section 2, if the 
\oxysix-rich gas is beyond the {\it{IRAS}} cirrus patch, then $a$ 
is likely between
0.4 and 1.0.

Assuming that $j_\nu / \kappa_\nu$ is approximately constant
%and equal to $j_\nu_0 / \kappa_\nu_o$ 
over the
range of $\nu$ where $\tau_\nu \ne 0$, 
the integrated intrinsic intensity is:
\begin{equation}
I_i = \frac{j_{\nu_o}}{\kappa_{\nu_o}} 
\int_0^\infty \tau_\nu d\nu
%= 2 \int_0^\infty \frac{j_\nu}{\kappa_\nu} d\nu \times \int_0^\infty 
%N_j \frac{\sqrt{\pi} e^2 f_{jk} \lambda}{m_e c b} 
%(1 - \frac{b_k}{b_j} e^{-h\nu/{\rm{k}}T}) e^{-(\frac{\lambda 
%\Delta \nu}{b})^2}
%d(\Delta\nu).
= \frac{2 j_{\nu_o}}{\kappa_{\nu_o}} \int_0^\infty 
\tau_o e^{-(\frac{\lambda \Delta \nu}{b})^2}
d(\Delta\nu)
= 
%\int_0^\infty \frac{j_\nu}{\kappa_\nu} d\nu \times N_j \frac{\pi e^2 
%f_{jk}}{m_e c} 
%(1 - \frac{b_k}{b_j}e^{-h\nu/{\rm{k}}T}) = 
\tau_o \frac{\sqrt{\pi} b}{\lambda} \frac{j_{\nu_o}}{\kappa_{\nu_o}}.
\end{equation}
%Thus,
%$\frac{I_o}{I_i} = \frac{A \lambda}{2 b} \sum_{n=1}^{\infty} 
%\frac{(-1)^{n-1} t^{n-1}}{n! \sqrt{n}}$.

We will use the ratio of equation (7) to equation (8) to calculate
the intrinsic intensities from the observed intensities.  
This requires knowing the optical depth
for each emission line: $\tau_{o(1032)}$ and $\tau_{o(1038)}$.
If the column density is known, then equation (4) can be used to
determine each $\tau_o$.  Otherwise, the optical depths
can be estimated from the ratio of the observed
intensities in the 1032 and 1038~\AA\ lines in the following manner.  
%Here we refine the relationship.
%The ratio of the observed 1032~\AA\ and 1038~\AA\ intensities is 
%related to $\tau_o$ through equation (7).  
The oscillator strength 
%and statistical weight of the upper energy level 
for the
1032~\AA\ transition is twice that for the 1038~\AA\ transition
and the ratio of $j_\nu / \kappa_\nu$ is given by Spitzer's
equation (3-31),
thus, 
$\tau_{o(1032)}$ is equal to $2 \tau_{o(1038)}$ and 
$\frac{j_{\nu_o(1032)}}{\lambda_{(1032)}\kappa_{\nu_o(1032)}}
/ \frac{j_{\nu_o(1038)}}{\lambda_{(1038)}\kappa_{\nu_o(1038)}}$
is equal to $(\frac{\lambda_{(1038)}}{\lambda_{(1032)}})^4$.
%and $j_{\nu(1032)}/\kappa_{\nu(1032)} 
%\approx j_{\nu(1038)}/\kappa_{\nu(1038)}$
The resulting ratio of equation (7) for the 1032~\AA\ line to 
equation (7) for the 1038~\AA\ line is:
\begin{equation}
%\frac{I_{{o(1032)}}}{I_{{o(1038)}}} = \frac{\sum_{n=1}^{\infty} 
%\frac{(-1)^{n-1} (2\tau_{o(1038)})^n}{n! \sqrt{n}}}
%{\sum_{n=1}^{\infty} \frac{(-1)^{n-1} (\tau_{o(1038)})^n}{n! \sqrt{n}}}.
\frac{I_{{o(1032)}}}{I_{{o(1038)}}} = 
\frac{\lambda_{(1038)}^4 F(2\tau_{o(1038)})}
{\lambda_{(1032)}^4 F(\tau_{o(1038)})}.
\end{equation}
%{\bf(I removed the factor of 2.  Is this right?)}.  
%I set up the summations in a computer program and the calculated
%ratio makes sense for the limiting cases of small and large taus.
This can be solved for $\tau_{o(1038)}$.

Next we determine the average electron density from the
intrinsic intensity and the column density.  
The following equation motivated our earlier
derivations of the intrinsic intensity and column density.
Shull \& Slavin (1994) present a relation for the 
average electron density of an emitting region in which the
kinetic temperature and emissivity per \oxysix\ ion are constant:
%\begin{equation}
\begin{eqnarray}
\lefteqn{\langle n_e \rangle  = 
	\frac{4 \pi}{\langle \sigma v \rangle_e}\; 
	\frac{(I_{i5000})}{N_{OVI}}  =  } \nonumber \\
	& & (0.00921\;{\rm cm}^{-3} )\; 
	\Biggl[ \frac{5}{\bar{\Omega}_{OVI}(T)} \Biggr] \;
	T_5^{1/2}\;\exp\;\biggl( \frac{1.392}{T_5} \biggr) \;
	\frac{I_{i5000}}{N_{14}} ,
\end{eqnarray}
%\end{equation}
where $\langle \sigma v \rangle_e$ is the electron-impact
excitation-rate coefficient for the 1032 and 1038~\AA\
lines combined, $\bar{\Omega}_{OVI}(T)$ the
Maxwellian-averaged collision strength for de-excitation of the \oxysix\
doublet, $T$ is the gas temperature, $T_5$ the gas temperature in $10^5$ K, 
$N_{14}$ the column density in units of $10^{14}$
cm$^{-2}$, and $I_{i5000}$ the
intrinsic \oxysix\ doublet intensity in units of 5000 photons cm$^{-2}$
s$^{-1}$ sr$^{-1}$.  
%By the intrinsic intensity, we mean the integral of the volumetric
%luminosity over the path length and divided by $4\pi$ steradians.
Shull \& Slavin present a
parameterization of the Maxwellian-averaged collision strength:
$\bar{\Omega}_{OVI}(T) = 6.43[0.5584 + (0.1363y - 0.1158y^2 +
0.279)E_1(y)e^y + 0.1158y]$, where $y = E_o/kT$ and $E_1(y)$ is the
first exponential integral, $\int_y^\infty \frac{e^{-x}}{x}dx$.  

The thermal pressure can also be calculated.   Assuming a cosmic abundance of
helium and complete ionization of both hydrogen and helium, 
\begin{equation}
P_{th}/k = 1.92 n_e T.  
\end{equation}
We expect the relevant temperatures to be near
$3.2 \times 10^5$ K because this is the 
temperature at which the fraction of 
oxygen atoms in the \oxysix\ ionization state is greatest
(Shapiro \& Moore 1976).  The \oxysix\ ions could exist within
plasmas of a range of temperatures.  For example,
the fraction of oxygen atoms in the \oxysix\ ionization state is at least
10$\%$ of its maximum value in collisional ionizational equilibrium
plasmas for temperatures between $2.2$ and $6.4\times 10^5$ K.
The electron density calculated from equation (10) 
varies little across this range.

The depth of the emitting region ($\Delta l$ or $\int dl$) is related to the column density
through $N_{OVI} = \int n_{OVI} dl$, where $n_{OVI}$ is the density
of \oxysix\ ions.  Thus, $\Delta l$ is approximately equal to 
$N_{OVI}/n_{OVI}$.
The volume density of \oxysix\ ions can be found from 
$n_{OVI} = n_e \times \frac{n_H}{n_e} \times \frac{n_{O}}{n_H} 
\times \frac{n_{OVI}}{n_{O}}$, where $n_e$, $n_H$, and $n_{O}$
are the densities of electrons, hydrogen atoms, and oxygen atoms respectively.
For a cosmic abundance of helium and complete ionization of both
hydrogen and helium, $n_e$ is equal to $1.2 n_H$.  Using the Grevesse
and Anders (1989) solar abundance of oxygen, 
$n_{O}$ is equal to $8.5 \times 10^{-4} n_H$. 
According to the Shapiro and Moore (1976) collisional ionizational
equilibrium curve for oxygen, the maximum value of $n_{OVI} / n_{O}$
is 0.26.  This occurs at $T = 3.2 \times 10^5$~K.
The resulting equation for $\Delta l$ is,
\begin{equation}
\Delta l = 5400 \times \frac{N_{OVI}}{n_e}, 
%{\rm{\ \ (in\ cm)\ \ or\ \ }}
%\Delta l = 1.75 \times 10^{-15} \times \frac{N_{OVI}}{n_e} 
%{\rm{\ \ (in\ pc)}}, 
\end{equation}
where $T$ is assumed to be $3.2 \times 10^5$~K, $N_{OVI}$ is
in cm$^{-2}$, $n_{e}$ is in cm$^{-3}$, and $\Delta l$ is in cm. 
Although most of the \oxysix\ ions probably reside within gas
having nearly the maximum value of $n_{OVI} / n_{O}$, some of the
ions may reside in hotter or cooler gas.  

\subsection{Calculated Values for the Physical Characteristics}

In this subsection, we use
equations (4), (7), (8), (9), (10), (11), and (12) to relate
the observed intensities,
optical depths, intrinsic
intensities, 
column densities, electron densities, thermal pressures, and
approximate emitting depth.  The results are presented
in Table 5 and Figure 5.
In case $\#1a$, we take $a = 1.0$ and use the observed intensities and
equation (9) to determine $\tau_{o(1038)}$, finding
%$\tau_{o(1038)} = 0.69_{-0.69{\rm{(r)}}}^{+1.06{\rm{(r)}}}$.
$\tau_{o(1038)} = 0.79_{-0.61{\rm{(r)}}}^{+1.38{\rm{(r)}}}$.
There is no systematic uncertainty in $\tau_o$ calculated from the
ratio of the observed intensities.  Then
$\tau_{o(1032)}$ is calculated from $\tau_{o(1038)}$.  
We use the optical depths, observed intensities, and the ratio
of equation (8) to equation (7) to calculate the intrinsic
intensities for the emission lines.  The
resulting intrinsic intensity of the doublet is 
%$6,700^{+16,800{\rm{(r)}}}_{-2,600{\rm{(r)}}}\pm950{\rm{(s)}}$
$7,100^{+19,100{\rm{(r)}}}_{-2,900{\rm{(r)}}}\pm1000{\rm{(s)}}$
photons s$^{-1}$ cm$^{-2}$ sr$^{-1}$.
We calculate the velocity parameter from the velocity FWHM in Table 2.
We use the velocity parameter and $\tau_o$ to calculate the 
column density in the ground state, finding 
%$N = 4.43_{-4.43{\rm{(r)}}}^{+6.82{\rm{(r)}}} \times 10^{14}$.
$N = 5.1_{-3.9{\rm{(r)}}}^{+8.9{\rm{(r)}}} \times 10^{14}$.
Note that due to the uncertainty in the observed line ratio,
the uncertainty in $N$ is very large.
We use the doublet intensity, column density, and equation
(10) to calculate the electron density for 
$T = 3.2 \times 10^5$~K, yielding
%$n_e = 0.0065^{+0.015{\rm{(r)}}}_{-0.0025{\rm{(r)}}}\pm0.0009{\rm{(s)}}$.
$n_e = 0.0060^{+0.016{\rm{(r)}}}_{-0.0024{\rm{(r)}}}\pm0.0008{\rm{(s)}}$.
As equation (10) and Figure 5 show, the electron
density varies little between $T = 2 \times 10^5$ and
$2 \times 10^6$~K.
We use equation (11) to calculate the thermal
pressure at this temperature, finding 
%$P_{th}/k = 4,000^{+9,400{\rm{(r)}}}_{-1,500{\rm{(r)}}}\pm560{\rm{(s)}}$.
$P_{th}/k = 3,700^{+9,900{\rm{(r)}}}_{-1,500{\rm{(r)}}}\pm520{\rm{(s)}}$.
Equation (12) is used to calculate the 
emitting depth if the gas is
at this temperature, finding $\Delta l = 
%3.7^{+10{\rm{(r)}}}_{-3.7{\rm{(r)}}}\pm{0.51}{\rm{(s)}} \times 
%10^{20}$ cm, or
%$120^{+330{\rm{(r)}}}_{-120{\rm{(r)}}}\pm{18}{\rm{(s)}}$ pc.
4.6^{+15{\rm{(r)}}}_{-4.0{\rm{(r)}}}\pm{0.64}{\rm{(s)}} \times 
10^{20}$ cm, or
$150^{+480{\rm{(r)}}}_{-130{\rm{(r)}}}\pm{21}{\rm{(s)}}$ pc.

This process is repeated for $a = 0.4$ in case $\#1b$.
Table 5 presents the results for each case.  In the table, $\Delta l$ is
presented in units of parsecs.
The calculated uncertainties are rather large for
some of the tabulated characteristics.
%For example, the derived column density minus $1 \sigma$ equals
%zero.  The derived value plus $1 \sigma$ can be taken as an
%upper limit.  
The main source of the propagated uncertainties in
cases $\#1b$ and $\#1b$ is the relatively large uncertainty in the
line ratio.
As a result, it is instructive to examine cases 
in which the column density (and hence the ratio of
observed intensities) is set by the assumption
of a particular geometry.  
In case $\#2$, the emission is assumed to originate in the Local Bubble.
For this case we take the absorption column density to be
$N_{OVI} \approx 1.6 \times 10^{13}$ cm$^{-2}$ (Shelton \& Cox
1994).  Only negligible extinction is expected, so $a$ is taken to be
1.0.
From the column density, we determine the optical
depths and the
ratio of intensities observed in the 1032~\AA\ and 1038~\AA\
emission lines.  The total observed intensity is assumed to be
conserved.  We then calculate the other quantities using
the same method as in cases $\#1a$ and $\#1b$.  The results are
presented in Table 5 and Figure 5.

In cases $\#3a$ and $\#3b$, the 
observed emission is assumed to originate in part
in the Local Bubble and in part
the Galactic thick disk and lower halo.  For these
cases, we take the absorbing column density to be 
$N_{OVI} = 2.0 \times 10^{14}$ cm$^{-2}$ from
Savage et al. (2000).  Because the previously discussed
cirrus may lie either closer than the thick disk/halo emission
region or beyond it,
the physical characteristics are calculated 
using $a=1.0$ (case $\#3a$) and $a=0.4$ (case $\#3b$).
The calculations follow the same method as in
case $\#2$ and are presented in Table 5 and Figure 5.
%Note that the calculations assume that the emissivity per
%\oxysix\ ion is constant along the line of sight.  Thus,
%it is assumed that the thermal pressure is constant.
Note that in this model, 
the Local Bubble, thick disk, and lower halo are assumed to
have the same thermal pressure.  This is the 
pressure calculated from
the equations and listed in Table 5.

\subsection{Caveats}

In this analysis, 
%By using equations (5) through (8), 
we have made the following assumptions, which we will discuss in 
more detail in the remainder of this section.
1.) The contribution of 1032 and 1038~\AA\ photons emitted by other
\oxysix\ ions in the ``cloud'' but
outside of the beam and scattered into the beam via interactions with
\oxysix\ ions
is assumed to be negligible when compared with the number of
such photons scattered out of the beam via such interactions.
2.) The contribution due to scattering of 
background ultraviolet continuum photons is assumed to be negligible.
3.) Losses due to other types of particles interspersed with
the \oxysix\ ions are assumed to be negligible (though the calculations
do consider an external region of extincting particles).
4.) The emissivity per \oxysix\ ion is assumed to be constant.
5.) Furthermore,
%Equations (5) through (8) do not require that the gas be in or near
%collisional ionizational equilibrium, but 
by using equations (10) and (11), we make the 
%for the thermal pressure and region depth 
assumption that the temperature 
and $n_{OVI}/n_{O}$ ratio are near their values at the peak of
the ion fraction curve constructed by Shapiro and Moore using 
collisional ionization equilibrium.  

If the first assumption is not valid 
then the intrinsic doublet intensities will be
less than the calculated values.
The degree of scattering into the beam is not known.  
%The limiting values are $0 < I_{si} < I_{so}$.  If $I_{si} \sim 0$, then
%scattering has no effect on our predicted intrinsic intensities.  
%If $I_{si} \sim I_{so}$, then the intrinsic intensities will 
%The least possible effect is if no 
%1032 or 1038~\AA\ photons produced by \oxysix\ ions are scattered into the
%beam.  In this case, the calculated intrinsic intensity would be correct.
If the number of photons produced by the
\oxysix\ ``cloud'' and scattered into the beam via interactions
with \oxysix\ ions compensates exactly
for the number produced by the \oxysix\ ``cloud'' and
scattered out via interactions
with \oxysix\ ions, then the intrinsic intensities would not
follow equation (7), but would be equal to the observed intensities after
correction for dust extinction.
%The anticipated greatest possible effect would be if the quantity scattered 
%into the beam
%compensates for the losses due to self absorption.  Then,
%in the absence of dust extinction, the intrinsic
%intensities would not follow equation (6) but would
%be equal to the observed intensities.
The results of such a scenario 
are within the error bars on our calculated doublet
intensities in each of our 
cases (i.e. $\#1a$, $\#1b$, $\#2$, $\#3a$, and $\#3b$).  
If the number of photons produced by the \oxysix\ ``cloud''
and scattered into the beam due to interactions
with \oxysix\ ions 
were to be even larger than
the number produced by the \oxysix\ ``cloud'' and
scattered out, then the true intrinsic intensity would
be less than the observed intensity after correction for dust
extinction.  In this scenario, we would expect
the 1032~\AA\ intensity to be at least twice as strong as the
1038~\AA\ intensity because \oxysix\ ions scatter 1032~\AA\ photons
more readily.  Such a scenario is marginally excluded because
the observed ratio of the 1032 and 1038~\AA\ intensities ($1.64 \pm 0.29(r)$)
is at least 1.24 sigma from the theoretical value ($>$2).
In cases $\#1a$ and $\#1b$, if some of the photons
scattered out of the beam have been compensated for by photons
scattered into the beam, then the true optical depths and column density
would be greater than the calculated values.  
For cases $\#2$, $\#3a$, and $\#3b$, 
the column densities were assumed and the optical depths were found from
them.    
If some of the light scattered out of the beam were replaced by light
scattered into the beam, then the ratio of the observed to the intrinsic
intensity would be less than that calculated from equations
(7) and (8) using the $\tau_o$ calculated
from the assumed column density.
In all cases, if the intensity scattered
into the beam is not negligible when compared with the intensity
scattered out of the beam, then the
true electron density and thermal pressure would be less than
the calculated values.

It is difficult to evaluate the second assumption,
%to calculate the 
%contribution due to scattering of stellar continuum photons, 
primarily because the geometric relationship
between the stellar continuum sources and \oxysix\ ions is not well known.
The following is an order of magnitude estimate.
We assume that the \oxysix\ ions on the line of sight are bathed
in the incident ultraviolet continuum field, whose strength
is:
%$I_{sci} = 2.27 \times 10^{-9}$ photons s$^{-1}$ cm$^{-2}$ sr$^{-1}$ Hz$^{-1}$
$I_{sci} = 4.38 \times 10^{-20}$ ergs s$^{-1}$ cm$^{-2}$ sr$^{-1}$ Hz$^{-1}$
(Mathis, Mezger, \& Panagia 1983).
The volumetric intensity absorbed by the \oxysix\ photons 
will be $I_{a \nu} = I_{sci} n_o \frac{\pi e^2 f}{m_e c}
\frac{\lambda}{b \sqrt{\pi}} e^{-(\lambda \Delta \nu / b)^2}$.
The ions will quickly re-emit this volumetric intensity, but
a fraction of the re-radiated
photons will arrive at the observer while 
other photons will be scattered out of the beam due to
absorption by intervening \oxysix\ ions or will be absorbed
by other intervening particles. 
The observed intensity ($I_{sco}$) can be found by solving
Spitzer's equation (3-1) or our equation (1), where $I_{a \nu}$ is 
substituted for
$j_\nu$ and where an extinction factor, $a$, is included.  
The result is
%The intensity arriving at the
%observer's location is 
%$dI_{sco,\nu} = a I_{sci} e^{-\tau_\nu '} d\tau_\nu '$.
%The integral with respect to $\tau'$ and $\nu$ is:
\begin{equation}
I_{sco} = \frac{a2b}{\lambda}{I_{sci}} 
F(\tau_o).
\end{equation}

We use equation (13) to create rough estimates of the 
possible contribution of scattered ultraviolet continuum 
photons to the observed intensity.  
The results are presented in the last column of Table 5.
The systematic uncertainties in $I_{sco}$ are bound to be large.
As a result, we are not prepared to make confident corrections to
the observed \oxysix\ intensities at this time.  However, it
is logical to recognize that if some scattered ultraviolet continuum
photons are being observed, then the true intrinsic intensity, 
electron density, thermal pressure, and emitting depth are less than
the values presented in Table 5.  In particular, if the 
estimated scattering is used, then
the doublet intensities, electron densities, and thermal pressures
in the most plausible cases, 
%cases $\#1a$, $\#1b$, $\#2$, $\#3a$, $\#3b$ would decrease
%by $78\%$, $32\%$, $78\%$, $4.0\%$, $43\%$, and $17\%$
cases $\#2$ and $\#3b$, would decrease from the quoted values 
by $4.0\%$ and $17\%$, respectively.  The emission depths would
increase by $4.0\%$ and $20\%$, respectively.

In the third assumption, there are no other absorbing particles
interspersed with the \oxysix\ ions.  If this assumption is false and if
the calculated extinction by the cirrus is not an overestimate,
then
the effective extinction would be greater than assumed.  Thus, the
intrinsic intensities, electron density, and thermal pressure
would be greater than the calculated values.  The emitting region
would be smaller than the calculated value.

In the fourth assumption, the emissivity per \oxysix\ ion is assumed to
be constant.  In case
$\#3$ where some
of the emission is from the Local Bubble and some is from the halo,
the thermal pressure varies along the path length.  Thus the
electron density and emissivity vary along the path length.  
The higher pressure gas in the Local Bubble would be more emissive
per ion than the lower pressure gas in the thick disk and halo.
As a result, in case $\#3$, the pressure listed in Table 5 would
lie between the pressure of the gas in the Local Bubble and the
pressure of the gas in the thick disk and halo.  This logic also
applies to case $\#1$.

If the assumptions regarding the temperature and ionization state of
the gas are invalid, then the electron density, thermal pressure, and
emitting depth are affected.  If the gas is near the specified temperatures
but out of collisional ionizational equilibrium, then the emitting
depths would be larger than calculated from equation (12).  
Alternatively, if the
fraction of oxygen ions in the \oxysix\ stage is near the peak in
the Shapiro and Moore curve, but the gas temperature is much
lower (or somewhat higher) than assumed, then the emissivity would be 
lower (or higher) than assumed and so the calculated electron
density would be higher (or lower) than assumed.

\subsection{Possible Locations of the Emitting Gas}

%The average densities and pressures estimated for the Local Bubble and
%Galactic halo cases above can be directly compared with other
%estimates physical conditions in these regimes and the 
%results used to constrain
%the most likely location for the emitting
%material.

In this subsection, we critique possible emission sources. 
We first examine the possibility that the observed emission
originates entirely in the Local Bubble.
{\it{ROSAT}} observations of soft X-ray emission from the Local Bubble 
suggest that the thermal pressure is $\sim15,000$ K \percc\ (Snowden et
al. 1998).  This value was found by fitting the soft X-ray spectrum with
model spectra for solar abundance, collisional ionization equilibrium plasma
and is appropriate for the $\sim10^6$~K, X-ray 
emitting plasma.  
%This pressure is significantly lower than the
%estimates made above assuming Local Bubble \oxysix\ absorption column
%densities.  
%Assuming that the gas being traced by the X-ray emitting gas
%is in pressure equilibrium with that traced by the ultraviolet emitting gas,
%the discrepancy between the Snowden et al. pressure
%estimate and the range of thermal pressures implied by the diffuse
%\oxysix\ emission suggests the latter does not predominately originate in
%the Local Bubble.  
%An alternate calculation yields the same conclusion. 
Cool clouds within the Local Bubble have been shown to have
lower thermal pressures (i.e. Jenkins (1998)).  We assume that the
difference between the thermal pressures of the hot and cool gas is
balanced by other forms of pressure, such as magnetic or turbulent pressure.
Hence, the thermal pressure in the cool gas does not constrain our
analysis.
The pressure of the soft X-ray emitting gas is within the
range of pressures found for case $\#2$.

If we assume that the \oxysix-bearing gas is in pressure balance with
the X-ray emitting gas and that the \oxysix\ ions predominately exist
within gas of temperatures between 2.2 and $6.4 \times 10^5 
{\rm{K}}$,
then the electron density is about
$\frac{15,000}{1.92 \times (2.2 {\rm{\ to\ }} 6.4 \times 10^5 
{\rm{K}})} = 0.012$ to $0.036$ cm$^{-3}$.  
Using equation (10) with these temperatures and electron densities and
the Shelton \& Cox (1994) value for the Local Bubble column 
yields an intrinsic doublet intensity between $\sim440$ and
$\sim1300$ photons cm$^{-2}$ s$^{-1}$ sr$^{-1}$.  
Not only is this range substantially lower than the intrinsic doublet
intensity calculated for case $\#2$, it is well below the
observed doublet intensity!
Scattered ultraviolet continuum photons probably cannot account for
the intensity discrepancy because the relatively small 
Local Bubble \oxysix\ column density is ineffective at scattering 
the assumed incident photon field.
Thus, we conclude that the Local Bubble, which is assumed to be
in pressure balance and collisional ionizational equilibrium, cannot
account for the bulk of the observed \oxysix\  photons.  An additional
source is needed.  
%This conclusion finds weak, independent confirmation
%in the observed ratio of the 1032~\AA\ and 1038~\AA\ intensities.
%If all of the observed photons were to have originated
%in a uniform, stationary sphere or shell of \oxysix\ ions associated with the
%Local Bubble, then the intensity of photons created by other \oxysix\
%ions in the bubble and scattered into the beam would have been important.
%The resulting line ratio would have been approximately 2.0
%rather than the observed 1.64$\pm$0.29(r).

The photons which do originate in the Local Bubble probably come
from a thin zone of hot gas.  
Using a Local Bubble pressure of 15,000 K cm$^{-3}$ and equations
(10), (11)
%, we find the electron density.
%Using the electron density and the Shapiro and Moore equilibrium plots
%of the ionization fraction as a function of temperature,
and (12),
we calculate the thickness of the \oxysix-bearing gas as a function
of temperature.
If the temperature is $T = 3.2 \times 10^5$~K, then
the fraction of oxygen atoms in the \oxysix\ ionizational state is 
at its maximum and $\Delta l = 1.1$~pc.  
If the temperature is 
$T = 2.4 \times 10^5$~K or $T = 6.4 \times 10^5$~K, where
the abundance of \oxysix\ ions is $10\%$ of its maximum value,
$\Delta l = 7.9$~pc or 23~pc, respectively.
Thus it is possible that the \oxysix\ ions reside in a thin shell
on the edge of the Local Bubble.  Such a structure is consistent
with the lack of \oxysix\ column density towards stars
located within the interior of the bubble (Oegerle et al. 2001).

%Observational estimates for the thermal pressure of the Galactic halo
%are somewhat sparse in the literature.  
%%Observations of the warm
%%ionized medium (WIM) have suggested average electron densities 
%%$n_e \sim 0.2$ \percc\ (Reynolds 1993), which can be coupled
%%with recent temperature estimates of $T\sim6,000 - 10,000$ K (Haffner
%%et al. 2000) to yield thermal pressures of order $P_{th}/k \sim 2400 -
%%4000$ K \percc\ in the WIM (assuming hydrogen is fully ionized and
%%helium is predominantly neutral; see Reynolds et al. 1998 and Reynolds
%%\& Tufte 1995).  
%Spitzer \& Fitzpatrick (1993) give cogent arguments regarding 
%the particle density and temperatures within the clouds observed
%towards the low-halo star HD 93521.  These authors suggest an average
%value of roughly $P_{th}/k \sim 1200$ K \percc\ along this sight line.
%Boulares \& Cox (1990) have given theoretical arguments suggesting a
%total (thermal$+$nonthermal) pressure 
%of $P/k \la 13,000$ K \percc\ might predominate at
%heights $z \la 2$ kpc from the midplane of the Galaxy.  Thus the range
%of observational and theoretical pressure determinations for material
%likely associated with the lower Galactic halo are in rough agreement
%with the values derived above given the observed \oxysix\ emission line
%strengths.
%
%Although the derived pressures seem to be consistent with those
%expected for the lower regions of the Galactic halo, 

If observed emission originates in the Local Bubble, thick disk,
and halo, 
then the derived pressure should be 
between the thermal pressures of the hot gases in these regions. 
As mentioned above, the thermal pressure
of the hot gas in the Local Bubble is thought to be about 
$\sim 15,000$ K cm$^{-3}$.  The thermal pressure in the 
thick disk and halo  
ranges from $\sim 0$ at
large distances from the Galactic midplane to 
$\sim 13,000$ K cm$^{-3}$ within 2 kpc of the plane in regions
where the total pressure is predominately in the form of thermal
pressure (Boulares \& Cox 1990).
The derived pressures for cases $\#3a$ and $\#3b$ are within
the allowed range.

From Table 5, depending on dust extinction,
the thickness of the emission zone would be
12 to 29 pc if $T = 3.2 \times 10^5$~K.
If the temperature is between $2.4$ and $6.4 \times 10^5$~K,
then the values would be larger by a factor of up to
7 and 20, respectively.
On a Galactic scale, these values are not large.  It is possible
that the \oxysix\ in the thick disk and lower halo
resides in midscale structures scattered about the thick disk and
lower halo.
%(such as high $z$ supernova remnants;
%Shelton, 1998, Slavin, McKee \& Hollenbach 2000).  
Such a scenario would 
be consistent with the large scale height (2.7~kpc) and high degree of
inhomogeneity found in the Savage et al. (2000) mini-survey.

We now examine the expected velocity for the pointing direction.
In general, the velocities of the highly-ionized
material in this part of the sky can be understood by assuming that
a thickened disk of
highly-ionized material roughly corotates with the underlying thin
disk of low-ionization gas (e.g., Lu et al. 1994; Sembach \& Savage
1992).   Our measured \oxysix\ velocities differ 
from the velocities expected from co-rotating
gas within 10 kpc of the plane by over $100$ km sec$^{-1}$. 
%In our direction Galactic rotation carries
%material within 10 kpc of the plane to negative LSR
%velocities, but the observed emission appears at positive LSR velocities.  
This discrepancy does not rule out the possibility that most of the
emitting material resides within the thick disk $/$ halo.  It is
quite plausible that the \oxysix\ resides within shock-heated
gas traveling over 
$100$ km sec$^{-1}$ relative to the speed expected from galactic
rotation.

A third hypothetical location for the 
observed material is  high- or
intermediate-velocity clouds and their environments, 
but this hypothesis appears to be contradicted
by the 21 cm observations.  Wakker et al. (2001)
have presented Parkes \hone\ 21 cm observations in the direction of the
star HD 3175, which lies $\sim14^\circ$ from the \fuse\ pointing,
showing an \hone\ intermediate-velocity cloud at $v_{\rm LSR} \approx
+73$ km s$^{-1}$ with $N(\mbox{H I}) \approx 10^{19}$ cm$^{-2}$.
%While speculative, we note that the emission could be associated with
%such intermediate-velocity gas in the southern hemisphere.
However, the \hone\ spectra from
the southern \hone\ 21 cm emission survey of the Instituto Argentino de
Radioastronom\'{\i}a (Arnal et al. 2000; see also Morras et al. 2000)
show no evidence
for an \hone\ intermediate-velocity cloud over the velocity range $v_{\rm
LSR} \approx +50$ to $+90$ km s$^{-1}$ within $\sim1\fdg5$ of of the
\fuse\ pointing (R. Morras 2000, private communication).  
%The lack of \hone\ emission near the \fuse\ pointing
%makes it difficult to conclude that the observed \oxysix\ emission is
%associated with intermediate-velocity \hone\ clouds in this general
%direction.
%
%If it is the case that the observed \oxysix\ emission arises from a source
%other than the Galactic halo, then the range of pressures and
%densities determined above are upper limits to those found in the
%Galactic halo (since we do not observe any emission at velocities
%appropriate for the halo).

\section{Summary}

We have observed emission by \oxysix\ ions in the Galaxy's general
diffuse ISM outside of supernova remnants or superbubbles.  
%Our
%observation of $l = 315.0^\circ$, $b = -41.3^\circ$ 
%and the {\it{HUT}} observation of $l = 218.2^\circ$, $b = 56.4^\circ$ 
%constitute the entire data set of such observations.
%The intensity level observed by FUSE is
%only $\frac{1}{4}$ that observed by HUT, indicating significant
%variability in the interstellar medium.
We have ruled out the possibilities that the observed photons
are stellar, instrumental or terrestrial airglow.
The observed intensities, velocities, and intrinsic widths 
are presented in Table~2.  
We have used the observed properties to estimate the
column density, intrinsic intensity, average
electron density, thermal pressure, and depth of the emitting gas.
The results are presented in Table~5.
% assuming the
%emission has its origins in the Local Bubble or low Galactic halo.
%If the luminous plasma lies beyond an observed patch of FIR cirrus,
%then a simple scaling should be applied to the intensities, and
%implied average electron density, and thermal pressure.
%The only phenomenological
%assumption in our calculations of the deduced quantities is that the
%luminosity per \oxysix\ ion is constant along the path length.  We did not
%need to assume collisional ionization equilibrium within the plasma.

The doublet intensity is larger than
expected for emission by the (assumed quiescent) Local Bubble alone.
Thus, we surmise that some of the observed intensity must
arise in the Galactic thick disk or lower halo.  There are no
\hone\ clouds in the southern \hone\ 21 cm emission survey of the 
Instituto Argentino de
Radioastronom\'{\i}a near this direction at similar velocities.
Thus,
high velocity clouds are probably not the source of the observed emission.
The emitting material appears to be confined to relatively shallow
regions such as the edge of the Local Bubble or 10 to 600 pc thick 
structures in the Galactic halo.

%Because \oxysix\ is the most efficient coolant of $\sim10^5$ to $\sim10^6$~K
%gas, the intrinsic intensities provide rough estimates of the cooling rate 
%of the hot interstellar plasma.  
%%The extreme limits are
%%about 4,700 to 44,000 photons cm$^{-2}$ s$^{-1}$ sr$^{-1}$, corresponding
%%to about $9.0 \times 10^{-8}$ to $8.5 \times 10^{-7}$ 
%%ergs cm$^{-2}$ s$^{-1}$ sr$^{-1}$.  
%In addition,
%the deduced quantities provide a constraint on
%phenomenological models for the hot gas.
%, especially when they
%are combined with the results from other searches for \oxysix\ emission,
%the growing catalog of \oxysix\ column density observations, and soft
%X-ray data sets.

%\pagebreak

\acknowledgements

We would like to thank K. D. Kuntz for reporting the $\frac{1}{4}$ keV
surface brightness for this part of the sky, Steve Snowden for
providing the {\it{ROSAT}} All Sky Survey and DIRBE-corrected {\it{IRAS}} 
maps of
the southern Galactic hemisphere, Jeff Valenti for advising on $H_2$
fluorescence, James Lauroesch for sharing his expertise on ultraviolet
emission lines, Riccardo Morras for kindly providing us with \hone\ spectra in
the vicinity of the \fuse\ pointing,  Shauna Sallmen for checking some
of the analysis,  
Tim Heckman for conversations about scattering of stellar
continuum photons, John Raymond and Dick Edgar for conversations about
radiative transfer, our referee, Wilt
Sanders, for his dedication in carefully reviewing the manuscript,
and members of the larger {\it{FUSE}} team for
acquiring and assisting in processing the data.  This article uses
data obtained during the In Orbit Checkout phase of the NASA-CNES-CSA
{\it{FUSE}} mission.  {\it{FUSE}} is operated by the Johns Hopkins
University.  U.S. {\it{FUSE}} PI-Team participants have received
financial support under NASA contract NAS5-32985.  This work made use
of the NASA SkyView facility (http://skyview.gsfc.nasa.gov).
\\

\noindent
{\large{\bf{References}}}

\noindent
Abgrall, H., Roueff, E., Launay, F., Roncin, J. Y., \& Subtil, J. L.
1993a, \aaps, 101, 273\\
Abgrall, H., Roueff, E., Launay, F., Roncin, J. Y., \& Subtil, J. L.
1993b, \aaps, 101, 323\\
Arnal, E.M., Bajaja, E., Larrarte, J.J., Morras, R., \& P\"{o}ppel, W.G.L.
	2000, \aaps, 142, 35\\
%
%Black, J. H. 1987, in Interstellar Processes, ed. D. J. Hollenbach \& H. A. 
%Thronson (Dordrecht: Reidel), 731\\
%
Boulares, A., \& Cox, D. P. 1990, ApJ, 365, 544\\
Breitschwerdt, D., Egger, R., Freyberg, M. J., Frisch, P. C., 
\& Vallerga, J. V. \\
\hspace*{0.6cm}1996, Space. Sci. Rev., 78, 183\\
Burrows, D. N., \& Mendenhall, J. A. 1991, Nature, 351, 629\\
Cox, D. P., \& Reynolds, R. J. 1987, ARAA, 25, 303\\
Curdt, W., Feldman, U., Laming, J. M., Wilhelm, K., Schuele, U., 
\& Lemaire, P. \\
\hspace*{0.6cm}1997, \aaps, 126, 281\\
%replaced Sch\"{u}le with Schuele to match ADS listing. 
%
Danly, L. \& Kuntz, K.D. 1993, in ``Star Formation, Galaxies, and
the Interstellar Medium",  \\
\hspace*{0.6cm}ed. J. J. Franco (New York:  Cambridge University Press), 86\\
Dickey, J.M. \& Lockman, F.J. 1990, ARAA, 28, 215\\
Diplas, A. \& Savage, B. D. 1994, ApJ, 427, 274\\
Dixon, W. V., Davidsen, A. F., \& Ferguson, H. C. 1996, ApJ, 465, 288\\
Dixon, W. V., Sallmen, S., Hurwitz, M., \& Lieu, R. 2001, in preparation\\
Edelstein, J., \& Bowyer, S. 1993, Adv. Space Res., 13, 307\\
Edelstein, J., Bowyer, C. S., Korpela, E., Lampton, M., Trapero, J., 
Gomez, J. F., \\
\hspace*{0.6cm}Morales, C., \& Orozco, V. 1999, BAAS, 195, 5302\\
%
%Friedman, S. D., {\it{et al}} 2000, ApJ, this volume\\
%
Fitzpatrick, E. L. 1999, PASP, 111, 63\\
%
%Fitzpatrick, E. L. \& Massa, D. 1988, ApJ, 328, 734\\
%
Fried, P. M., Nousek, J. A., Sanders, W. T., \& Kraushaar, W. L.
1980, ApJ, \\
\hspace*{0.6cm}242, 987\\
Frisch, P. C. 1995, Space Sci. Rev., 72, 499\\
Grevesse, N., \& Anders, E. 1989, in AIP Conf. Proc. 183, 
Cosmic Abundances\\ 
\hspace*{0.6cm}of Matter, ed. C. J. Waddington (New York: AIP), 1\\
%
%Haffner, L.M., Reynolds, R.J., \& Tufte, S.L. 1999, ApJ, 523, 223\\
%
Hog E., Fabricius C., Makarov V.V., Urban S., Corbin T., Wycoff G., 
Bastian U., \\
\hspace*{0.6cm}Schwekendiek P., \& Wicenec A. 2000, A\&A, 
355, L27\\
Holberg, J. B. 1986, ApJ, 311, 969\\
%
%Horne, K. 1986, PASP, 98, 609\\
%
%Houk, N., A.P. 1975, University of Michigan Catalogue of Two-dimensional Spectral Types for the HD Stars, Vol. 1 (Ann Arbor: Univ. Michigan, Department of Astronomy)
Houk, N. \& Cowley, A.P. 1975, Michigan Spectral Catalog
(Ann Arbor: University of\\
\hspace*{0.6cm} Michigan, Department of Astronomy) \\
Jenkins, E. B. 1996, ApJ, 471, 292\\
%
%Jenkins, E. B. 1978, ApJ, 219, 845\\
%
Jenkins, E. B. 1998, in IAU Colloquium Proc. 166, 
Lecture Notes in Physics, vol. 506, \\
\hspace*{0.6cm} The Local Bubble and Beyond,
 Proceedings of the IAU Colloquium No. 166, ed. D. 
\hspace*{0.6cm} Breitschwerdt, M. J. Freyberg, and 
J. Truemper, (Springer-Verlag), 33\\
Korpela, E. J., Bowyer, S. \& Edelstein, J. 1998, ApJ 495, 317\\
Kuntz, K. D. 2000, personal communication\\
Lallement, R., \& Bertin, P. 1992, A\&A, 266, 479\\
Lallement, R., Ferlet, R., Lagrange, A. M., Lemoine, M., \& 
Vidal-Madjar, A. \\
\hspace*{0.6cm}1995, A\&A, 304, 461\\
Linsky, J. L., Redfield, S., Wood, B. E., \&  Piskunov, N. 2000, ApJ, \\
\hspace*{0.6cm}528, 756\\
Lockman, F. J., Hobbs, L. M., \& Shull, J. M. 1986, ApJ, 301, 380\\
%
%Long, K. S., Blair, W. P., Vancura, O., Bowers, C., Davidsen, A. F., \&\\
%\hspace*{0.6cm}Raymond, J. C. 1992, ApJ, 400, 214\\
%
Lu, L., Savage, B.D., \& Sembach, K.R. 1994, ApJ, 437, 119L\\
McCammon, D., Burrows, D. N., Sanders, W. T., \& Kraushaar, W. L.
1983, \\
\hspace*{0.6cm}ApJ, 269, 107\\
Martin, C. \& Bowyer, S. 1990, ApJ, 350, 242\\
Mathis, J. S., Mezger, P. G., \& Panagia, N. 1983,
A \& A, 128, 212\\
%
%Moos, H. W., et al. 2000, ApJ 538, 1\\
%
Morras, R. 2000, private communication \\
Morras, R., Bajaja, E., Arnal, E.M., P\"{o}ppel, W.G.L. 2000, \aaps,
	142, 25\\
Morton, D. C. 1991, ApJS, 77, 119 \\
Morton, D. C., 
%``Atomic Data for Resonance Absorption Lines.  
%III.  Wavelengths Longward of the Lyman Limit for the Elements hydrogen 
%to Gallium'', 
2001, in preparation\\
Morton, D. C., \& Dinerstein, H. L. 1976, ApJ, 204, 1\\
Murthy, J., Henry, R. C., Shelton, R. L., \& Holberg, J. B. 2001, ApJ,\\
\hspace*{0.6cm} submitted\\
Oegerle, W., et al. 2001, in preparation\\
Oegerle, W., Murphy, E., \& Kriss, J. 2000, {\it{The FUSE DATA Handbook}},\\
\hspace*{0.6cm}http://fuse.pha.jhu.edu/analysis/dhbook.html\\
Redfield, S., \& Linsky, J. L. 2000, ApJ, 534, 825\\
Reynolds, R. J. 1993, in AIP Conf. Proc. 278, Back to the Galaxy, ed. \\
\hspace*{0.6cm}S. S. Holt \& F. Verter (New York: AIP), 156\\
%
%Reynolds, R.J., Hausen, N.R., Tufte, S.L., \& Haffner, L.M. 
%	1998, \apj, 494, L99\\
%
%Reynolds, R.J., \& Tufte, S.L. 1995, 439, L17\\
%
Sahnow, D. J., et al. 2000, ApJ, 538, L7\\
Sanders, W. T., Kraushaar, W. L., Nousek, J. A., \& Fried, P. M.
1977, ApJ,\\
\hspace*{0.6cm} 217, L87\\
Savage, B. D., et al. 2000, ApJ, 538, L27\\
Savage, B. D., Sembach, K. R., \& Lu, L. 1997, AJ 113, 2158\\
Sembach, K.R., \& Savage, B.D. 1992, \apjs, 83, 147\\
Sfeir, D. M., Lallement, R., Crifo, F., \& Welsh, B. Y.,
1999, A \& A, 346, 785\\
Shapiro, P.R., \& Moore, R. T. 1976, ApJ, 207, 460\\
Shelton, R. L., \& Cox, D. P. 1994, ApJ, 434, 599\\
%
%Shelton, R. L. 1998, ApJ, 504, 785\\
%
Shull, J. M., \& Slavin, J. D. 1994, ApJ, 427, 784\\
%
%Slavin, J. D., McKee, C.F., \& Hollenbach, D.J. 2000, ApJ, 541, 218\\
%
Snowden, S. L., Cox, D. P., McCammon, D., \& Sanders, W. T. 1990,
ApJ, \\
\hspace*{0.6cm}354, 211\\ 
Snowden, S. L., Egger, R., Finkbeiner, D. P., Freyberg, M. J., \& \\
\hspace*{0.6cm}Plucinsky, P. P.,
1998, ApJ, 493, 715\\
Snowden, S. L., Egger, R., Freyberg, M. J., McCammon, D., Plucinsky, P. P., \\
\hspace*{0.6cm}Sanders, W. T., Schmitt, J. H. M.M., 
Tr\"{u}mper, J., \& Voges, W. 1997, \\
\hspace*{0.6cm}ApJ, 485, 125\\
Snowden, S. L., Freyberg, M. J., Kuntz, K. D., and Sanders, W. T.,
2000, ApJS, 128, 171\\
Snowden, S. L., Mebold, U., Hirth, W., Herbstmeier, U., \& \\
\hspace*{0.6cm}Schmitt, J. H. M. 1991, Science, 252, 1529\\
Spitzer, L. 1978, Physical Processes in the Interstellar Medium 
(New York: Wiley)\\
%
%Spitzer, L., \& Fitzpatrick, E.L. 1993, \apj, 409, 299\\
%
%Sutherland, R.S., \& Dopita, M.A. 1993, \apjs, 88, 253\\
%
Sternberg, A. 1989, ApJ, 347, 863\\
Wakker, B.P., Kalberla, P.M.W., van Woerden, H., de Boer, K.S., \& 
	Putman, M.E. 2001, \\
\hspace*{0.6cm}\apjs, submitted\\
Wakker, B. P., \& van Woerden, H. 1997, AARA, 35, 217\\
Warwick, R. S., Barber, C. R., Hodgkin, S. T., \& Pye, J. P.
1993, MNRAS, 262, 289\\
Welsh, B. Y., Sfeir, D. M., Sirk, M. M., Lallement, R 1999, 
A\&A, 352, 308\\
%
%Welty, D. E., Frisch, P. C., Sonneborn, G., \& York, D. G. 1999, ApJ,
%512, 636\\
%
Wheelock et al. 1994, {\it{IRAS}} Sky Survey Atlas:  Explanatory Supplement
(Pasadena: JPL94-11)\\

%\newpage
%{\large\bf Figure Captions:}
%{\large\bf Figures:}

\vspace{-1in}
\begin{figure}
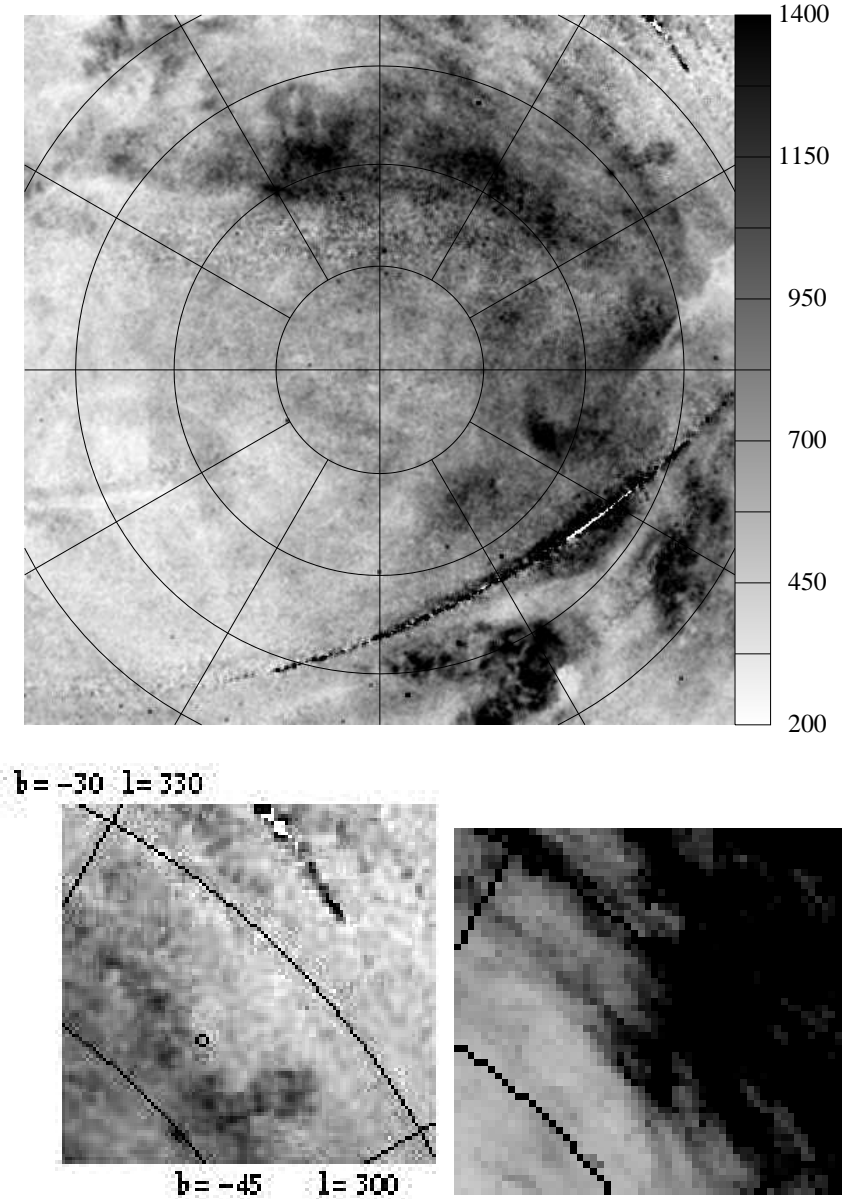

%%{\vspace{-2in}\epsscale{0.8}
%%\plotone{xraymap.ps}}
%%{\vspace{-1in}\epsscale{1.0}
%%\plottwo{xraysubmap.ps}{dirbeiraslinearbwsub.ps}}
%%\plotone{xraymap.ps}}
%{\vspace{-0.0in}\epsscale{0.7}
%\plotone{figure1a.epsi}}
%{\vspace{-0.0in}\epsscale{0.8}
%\plottwo{figure1b.epsi}{figure1c.epsi}}
\epsscale{0.65}
%\plotone{xraymap.epsi}\\
\plotone{figure1a.epsi}\\
\bigskip
\epsscale{0.342} %\epsscale{0.311}
%\plotone{xraysubmap.epsi}
\plotone{figure1b.epsi}
\epsscale{0.311} %\epsscale{0.40}
%\plotone{dirbeiraslinearsub.epsi}
\plotone{figure1c.epsi}
\caption{Top: The {\it{ROSAT}} $\frac{1}{4}$ keV surface brightness map of the
southern Galactic pole reproduced from Snowden et al. (1997), courtesy
of Steve Snowden.  The Galactic latitude is $-90^{\rm{o}}$ at the
center of the map and increases outward.  Lines of latitude are drawn
every 15$^\circ$.  The Galactic longitude is $0^\circ$ at the top of
the map.  Longitude increases counterclockwise, with meridians drawn
every $30^\circ$.  The units on the sidebar are $10^{-6}$ counts
s$^{-1}$ arcmin$^{-2}$.  The dark arc running across the lower right
quadrant indicates a region lacking sufficient {\it{ROSAT}} All Sky Survey
data.  The $\frac{1}{4}$ keV surface brightness for $l = 315^\circ$,
$l = -41^\circ$ is approximately average for the region of the
southern Galactic hemisphere away from the plane.  Bottom Left:
Enlargement of the upper right corner of the X-ray map.  A circle is
drawn around $l = 315^\circ$, $b = -41^\circ$.  Bottom Right:
DIRBE-corrected {\it{IRAS}} 100 $\mu$m map, reproduced from Snowden et
al. 1997.}
\end{figure}

\noindent
\begin{figure}
\epsscale{0.5}
\plotone{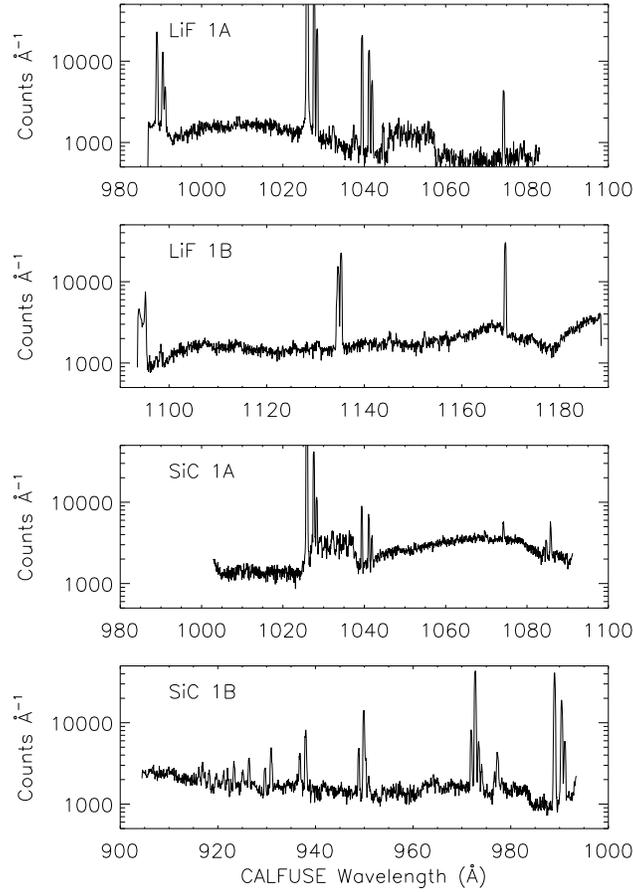}
%\epsscale{0.5}
%\plotone{emissionlif1b.ps}
%\plotone{emissionlif1a.ps}
%\plotone{emissionsic1a.ps}
%\plotone{emissionsic1b.ps}
\caption{
The counts spectra from the $l = 315^\circ$, $b = -41.3^\circ$ 
pointing.  Each spectrum is logorithmically scaled in counts,
includes satellite day and night data, and
is binned by 11 pixels ($\sim0.075$~\AA\ for the LiF 1A).
The strong, narrow peaks
are terrestrial airglow emission lines.
The elevations in the LiF 1A count rate at short
wavelengths and the SiC 1A countrate at long wavelengths
are due to elevated scattered light levels.
The small elevations in the intensity
between $\sim$1045 and $\sim$1055~\AA\ in the LiF 1A spectrum
and between $\sim$1025 and $\sim$1035~\AA\ in the SiC 1A spectrum are due to
a localized band of scattered light cutting across the spectrum.}
%\label{totalspectrum}
\end{figure}

\noindent
\begin{figure}
\plotone{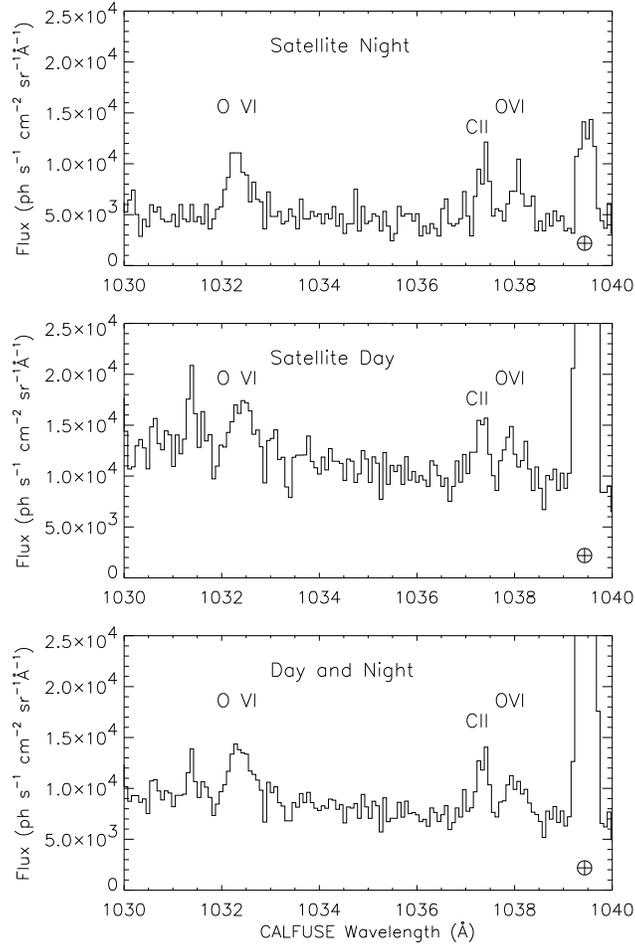}
\caption{
The LiF 1A spectrum, binned over 11 pixels (corresponding to 
0.075~\AA\ or about one fifth of the width of the large aperture) and
plotted relative to the uncorrected CALFUSE wavelengths.
%Note that unlike the \hone\ Ly $\beta$ and \oxyone\ airglow lines (at
%1025.7, 1027.4, 1028.2, 1039.2, 1040.0, 1040.9~\AA), 
%the \oxysix\ features persist in the night-only data,
%as a feature originating in the ISM must.
%An emission feature at 1037~\AA\ also persists in the night only data and has
%tentatively been identified as interstellar \cartwo\ 
%though H$_2$ fluorescence or
%absorption could modify its shape.  The emission peak at 1031~\AA\ 
%in the day and day$+$night spectra is absent in the night spectrum and
%therefore does not originate in the diffuse ISM
Top: spectrum from the satellite night portion of the data.  
Middle: spectrum from the satellite day portion of the data.  
Bottom:  spectrum from the day and night data.  
%The \hone\ and \oxyone\ atmospheric airglow emission lines 
%(at $\sim$1026, $\sim$1027, $\sim$1028, $\sim$1039, $\sim$1040, $\sim$1041,
%and $\sim$1042~\AA), are
The emission lines associated with the Earth's atmosphere
(each denoted by $\oplus$) are
quite strong in the total and day-only spectra,
but much weaker in the night-only spectra.  
The $\sim$1032 and $\sim$1038~\AA\ emission features of interstellar \oxysix\
have similar brightnesses in the night-only and the day-only spectra.
The 1037~\AA\ feature of interstellar \cartwo\
is also present in all spectra.
The 1031~\AA\ feature appearing in the total and
day-only spectra is absent in the night-only spectra and is not
statistically significant in the day-only data, thus is probably
not cosmic. 
%\label{ovidayvsnight}
}
\end{figure}

\begin{figure}
\plotone{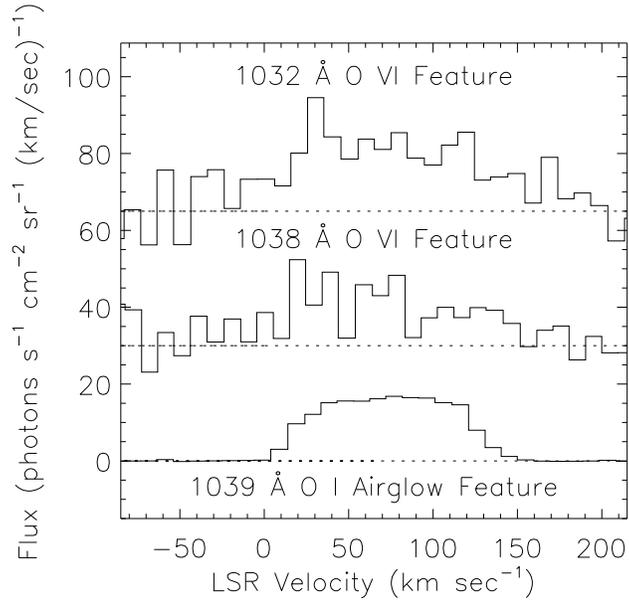}
\caption{
The 1032 and 1038~\protect\AA\ \protect\oxysix\ and 1039~\protect\AA\
\protect\oxyone\ profiles
versus LSR velocity and binned by 5 detector pixels.
Top:  The intensity in the \protect\oxysix\ 1032~\protect\AA\ emission feature.
For plotting purposes, the curve has been shifted upwards
by 65 photons s$\protect ^{-1}$ cm$\protect ^{-2}$ sr$\protect ^{-1}$ 
(km s$\protect ^{-1}$)$\protect ^{-1}$.
Middle:  Intensity in the \protect\oxysix\ 
1038~\protect\AA\ feature, shifted upwards 
by 30 photons s$\protect ^{-1}$ cm$\protect ^{-2}$ sr$\protect ^{-1}$ 
(km s$\protect ^{-1}$)$\protect ^{-1}$.
The 1032~\protect\AA\ and 1038~\protect\AA\ 
features have similar central velocities,
widths, and roughly similar envelopes. 
Bottom:  To illustrate that the instrument optics convolve the
intrinsic profile with a top hat function, 
the observed airglow \protect\oxyone\ 1039~\protect\AA\ profile, 
scaled by 0.3 and shifted to the right
by 65 km s$\protect ^{-1}$, is also shown.
%\label{velocity}
}
\end{figure}

\begin{figure}
\plotone{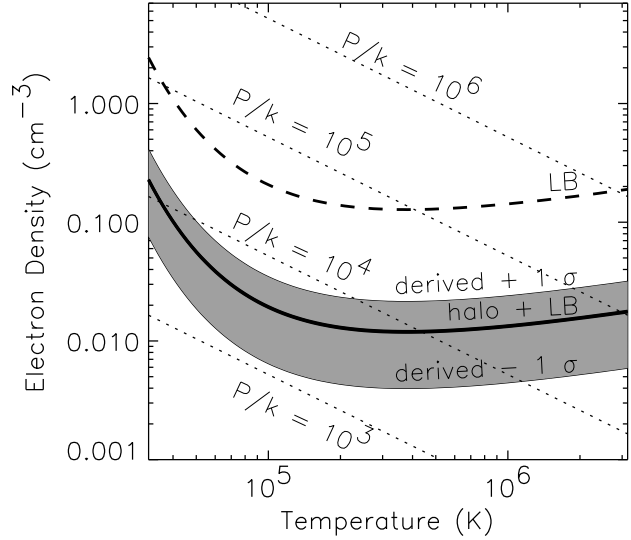}
%  \vspace{-1cm}
%%  \hspace*{-2.0cm}\psfig{file=vandensity3.ps,width=4.0in}
%  \hspace*{1.0cm}\psfig{file=electrondensity.ps,width=10cm}
%  \vspace{-0.5cm}
\caption{
The electron density $n_e$ as a function of 
temperature for an \oxysix\ plasma, assuming 
%an intrinsic $\lambda 1032, 1038$ surface brightness of 
%$I_{5000} =1.47$ and 
values calculated directly from the data (case $\#1a$, shaded region),
values calculated from the data and
an assumed \oxysix\ Local Bubble column density
(case $\#2$, dashed line),
and values calculated from the data 
and an assumed Local Bubble plus thick disk and lower halo
column density
(case $\#3a$, thick solid line).
%We reproduce
%the results of Shull \& Slavin (1994; dashed curve) for the
%\carfour-rich halo gas observed by Martin \& Bowyer (1990).
Overplotted are lines of constant thermal pressure ($P_{th}/k = 1.92 n_e T$,
dotted lines).
%\label{density}
}
\end{figure}

\newpage

\begin{table}
\begin{center}
\caption{Stars within 60 arcmin of $l = 315.0^\circ$, $b = -41.3^\circ$}
\vspace{0.5cm}
\begin{tabular}{cccccc}
\hline \hline 
Star   & Ang. Sep. & Sp. Class & Phot. Dist. & Par. Dist. & E(B-V) \\ 
       & (arcmin)       &           & (pc)         & (pc)       & \\ 
\hline
\ \ HD214691\ \ & 37.0 & G5V      & 40  & 49$\stackrel{+2}{_{-2}}$     & 0.05  \\
HD213591& 51.0 & G0V      & 53  & 47$\stackrel{+2}{_{-2}}$     & -0.02 \\
HD213928& 53.0 & F7/F8V   & 65  & 77$\stackrel{+5}{_{-5}}$     & 0.01  \\
HD214778& 40.0 & G2V      & 97  & 249$\stackrel{+119}{_{-61}}$ & -0.05 \\
HD214522& 59.0 & F8/G0V   & 134 &                           & -0.02 \\
HD213742& 41.0 & K1III    & 167 & 175$\stackrel{+22}{_{-17}}$  & 0.07  \\
HD216379& 58.0 & K2III    & 381 & 353$\stackrel{+190}{_{-92}}$ & 0.07  \\
HD215970& 39.0 & \ G8/K0III\  & 741 &                           & 0.02  \\ \hline
\end{tabular}
\end{center}
%\caption{Stars within 60 arcmin of $l = 315.0^\circ$, $b = -41.3^\circ$
%the pointing direction, their 
%identifications, angular separations from the pointing direction, spectral
%classes, distances from Earth (calculated from the photometric data),
%distances from Earth (calculated from the parallax data), and E(B-V)'s.
%}
\label{ebv}
\end{table}

\clearpage

\begin{deluxetable}{lcc}
%\rotate
\tablewidth{0pt}
\tablecaption{\oxysix\ Observations}
\tablehead{
\colhead{Characteristic}     
& \colhead{O~VI\ \ 1032 \AA\ feature} 
& \colhead{O~VI\ \ 1038 \AA\ feature}}
\startdata
Intensity (photons s$^{-1}$ cm$^{-2}$ sr$^{-1}$) & 2930  & 1790 \\
Intensity (photons s$^{-1}$ cm$^{-2}$ sr$^{-1}$) & 2930  & 1790 \\
$\sigma$ due to random uncertainties (photons s$^{-1}$ cm$^{-2}$ sr$^{-1}$)  & 290   & 260 \\
$\sigma$ due to systematic uncertainties (photons s$^{-1}$ cm$^{-2}$ sr$^{-1}$)  & 410   & 250 \\
Center LSR Velocity (km s$^{-1}$)                & $+70$ & $+60$ \\
Intrinsic FWHM (km s$^{-1}$) 			 & $\sim110$ & $\sim110$ \\ 
Kinetic Temperature (K) & $\la 4 \times 10^6$ & $\la 4 \times10^6$ \\ 
\enddata
\label{ovimeasurements}
\end{deluxetable}

%\begin{table}
%\begin{center}
%\caption{Characteristics of the \oxysix\ Emission Features}
%\vspace{0.5cm}
%\begin{tabular}{lcc}
%\hline \hline 
%Characteristic   & \oxysix\ \ 1032~\AA\ feature & \oxysix\\ 1038~\AA\ feature  \\ 
%\hline
%Intensity (photons s$^{-1}$ cm$^{-2}$ sr$^{-1}$) & 2930  & 1790 \\
%$\sigma$ due to random uncertainties (photons s$^{-1}$ cm$^{-2}$ sr$^{-1}$)  & 290   & 260 \\
%$\sigma$ due to systematic uncertainties (photons s$^{-1}$ cm$^{-2}$ sr$^{-1}$)  & 410   & 250 \\
%Center LSR Velocity (km s$^{-1}$)                & $+70$ & $+60$ \\
%Intrinsic FWHM (km s$^{-1}$) 			 & $\sim110$ & $\sim110$ \\ 
%Kinetic Temperature (K) & $\la 4 \times 10^6$ & $\la 4 \times10^6$ \\ 
%\hline
%\end{tabular}
%\end{center}
%\label{ovimeasurements}
%\end{table}

\clearpage

\begin{table}
\begin{center}
\caption{Characteristics of the \cartwo\ Emission Feature}
\vspace{0.5cm}
\begin{tabular}{lc}
\hline \hline 
Characteristic   & \cartwo\ 1037~\AA\ feature \\ 
\hline
Intensity (photons s$^{-1}$ cm$^{-2}$ sr$^{-1}$) & 1990 \\
$\sigma$ due to random uncertainties (photons s$^{-1}$ cm$^{-2}$ sr$^{-1}$) & 260 \\
$\sigma$ due to systematic uncertainties (photons s$^{-1}$ cm$^{-2}$ sr$^{-1}$)  & 280 \\
Center LSR Velocity (km s$^{-1}$) & $+$40 \\
\end{tabular}
\end{center}
%\caption{
%Characteristics of the \cartwo\ emission feature.
%Measurements for the \cartwo\ emission feature, including
%the intensity, sigma, and results of Gaussian fits.  The breadth of the
%{\it{FUSE}} large aperture limits the minimum Gaussian FWHM to about the
%value observed for the \cartwo\ emission feature.  The
%\cartwo\ emission line is unresolved.
%Thus the true
%emission line must be narrower than the observed width.
%}
\label{ciimeasurements}
\end{table}

\clearpage

\begin{table}
\begin{center}
\caption{3$\sigma$ Upper Limits for Other Transitions}
\vspace{0.5cm}
\begin{tabular}{ccc}
\hline \hline 
Species & Rest Wavelength & 3$\sigma$ Upper Limit$^{a}$\\ 
        & (\AA)           & (photons s$^{-1}$ cm$^{-2}$ sr$^{-1}$) \\ \hline
S~{\footnotesize{VI}}   & 933.378  & 3600 \\
S~{\footnotesize{VI}}   & 944.523  & 3600 \\
C~{\footnotesize{III}}  & 977.020  & 2600 \\
S~{\footnotesize{III}}  & 1015.502 & 780  \\
Si~{\footnotesize{II}}  & 1023.700 & 390  \\
%P~II   & 1156.970 & 3x690 but this is changed by the wormless effective
% area curve. \\ \hline
S~{\footnotesize{IV}}   & 1062.664 & 350  \\
C~{\footnotesize{I}}    & 1122.260 & 580  \\
Fe~{\footnotesize{III}} & 1122.524 & 580  \\ \hline
\end{tabular}
\end{center}
%\tablenottext{a}{As described in the text,
%these measurements are also subject to the systematic
%uncertianties of $14\%$.}
Note -- $a$:\ \ \ As described in the text,
these measurements are also subject to the systematic
uncertianties of $14\%$.
\label{otherspecies}
\end{table}

\clearpage

\begin{table}
\vspace{-1.5in}\hspace{-3in}\epsscale{1.0}\plotone{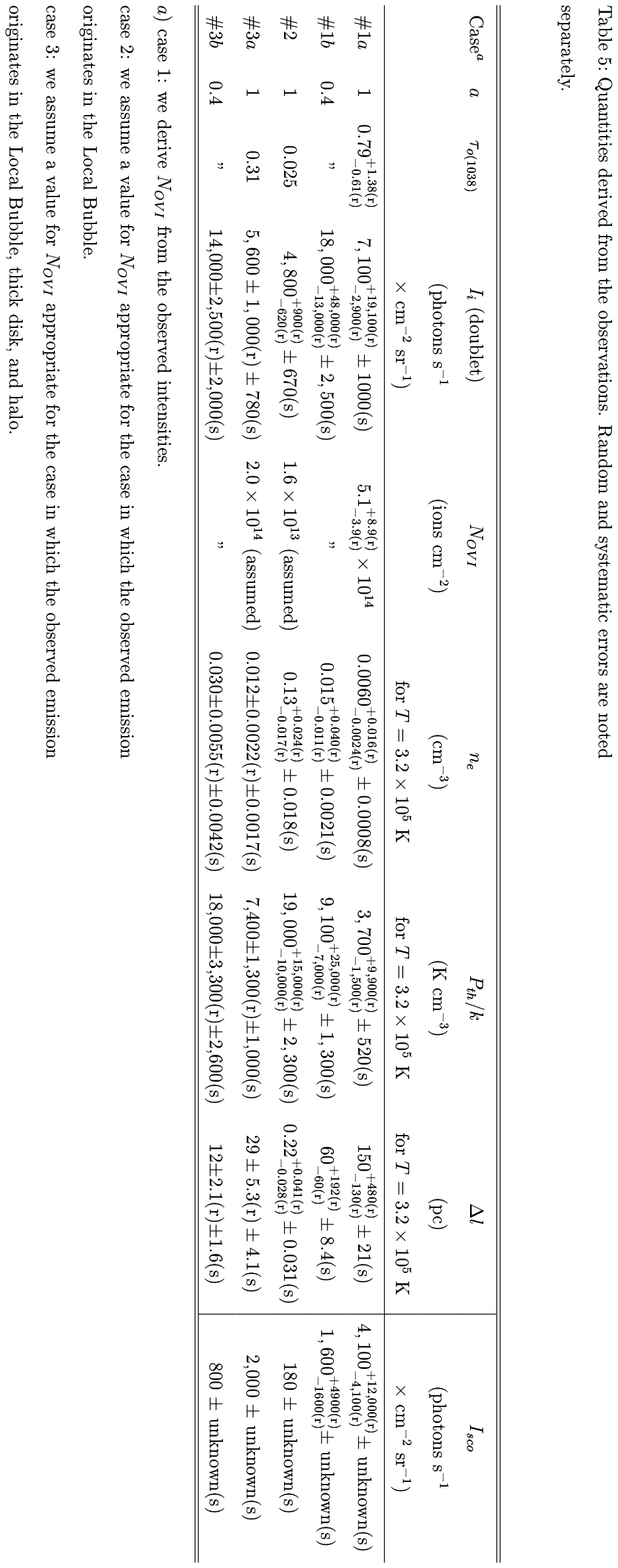}
\end{table}

\end{document}